\pdfoutput=1
\documentclass[prd,superscriptaddress,11pt,nofootinbib]{revtex4-1}
\usepackage{amsmath}    
\usepackage[english]{babel}
\usepackage[utf8]{inputenc}
\usepackage{graphicx}   
\usepackage{slashed}
\usepackage{epstopdf}
\usepackage{verbatim}   
\usepackage{color}      
\usepackage{subcaption}
\usepackage{multirow}
\usepackage{hyperref}   
\usepackage{float}
\restylefloat{table}
\usepackage{bm}
\usepackage[normalem]{ulem}
\newcommand{\be}{\begin{eqnarray}}
\newcommand{\ee}{\end{eqnarray}}

\def\addresses#1#2{\hbox to \hsize{\@tablebox{#1}\hfil\@tablebox{#2}}}
\def\@tablebox#1{\vtop{\hsize=5in \begin{flushleft} #1 \end{flushleft}}}

\def\beq{\begin{equation}}
\def\eeq{\end{equation}}
\def\bit{\begin{itemize}}
\def\eit{\end{itemize}}
\def\beqa{\begin{eqnarray}}
\def\eeqa{\end{eqnarray}}
\def\bray{\begin{array}}
\def\eray{\end{array}}

\def\n{\nonumber}

\def\zb{\bar z}

\definecolor{orange}{rgb}{1,0.5,0}
\definecolor{blue}{rgb}{0,0,1}

\begin{document}
\title{{\bf Light Dark Matter Showering under Broken  Dark $U(1)$ -- Revisited }}

\author{Junmou Chen}
\email{jmchen@kias.re.kr}
\affiliation{School of Physics, Korean Institute for Advanced Study, Seoul, 02455, Korea}

\author{Pyungwon Ko}
\email{pko@kias.re.kr}
\affiliation{School of Physics, Korean Institute for Advanced Study, Seoul, 02455, Korea}
\affiliation{Quantum Universe Center, Korean Institute for Advanced
Study, Seoul, 02455, Korea}

\author{Hsiang-nan Li}
\email{hnli@phys.sinica.edu.tw}
\affiliation{Institute of Physics, Academia Sinica, Taipei, Taiwan 115, Republic of China}

\author{Jinmian Li}
\email{jmli@kias.re.kr}
\affiliation{School of Physics, Korean Institute for Advanced Study, Seoul, 02455, Korea}

\author{Hiroshi Yokoya}
\email{hyokoya@kias.re.kr}
\affiliation{Quantum Universe Center, Korean Institute for Advanced
Study, Seoul, 02455, Korea}

\begin{abstract}
It was proposed recently that different chiralities of the dark matter 
(DM) fermion under a broken dark U(1) gauge group 
can lead to distinguishable signatures at the LHC through
shower patterns, which may reveal the mass origin
of the dark sector. We study this subject further by examining the 
dark shower of two simplified models, the
dubbed Chiral Model and the Vector Model.  We derive a more 
complete set of collinear splitting functions with power corrections, 
specifying the helicities of the initial DM fermion and including the 
contribution from an extra degree of freedom, the dark Higgs boson. 
The dark shower is then implemented with these splitting functions, and 
the new features resulting from its correct modelling are emphasized. 
It is shown that the DM fermion chirality can be differentiated by measuring 
dark shower patterns, especially the DM jet energy profile, which is  
almost independent of the DM energy.
\end{abstract}


\maketitle

\section{Introduction}

The nature of dark matter~(DM) remains one of the most challenging puzzles in
modern physics.   One of the popular scenarios is that the DM is composed of
weakly interacting massive particles (WIMP)~\cite{Lee:1977ua}, as strongly
motivated by the supersymmetric framework~\cite{Jungman:1995df}.  Null results 
of direct detection and LHC search have highly constrained this scenario
in recent years.  Meanwhile, new evidences such as the positron
excess in cosmic ray spectra~\cite{Adriani:2008zr}, the tension between
the cold DM model and the small structure
observations of the universe~\cite{Spergel:1999mh}, etc.\ have led us to consider 
other options.  One possibility is that there exists a new interaction in the dark
sector~\cite{ArkaniHamed:2008qp,Alexander:2016aln,ArkaniHamed:2008qn,Cirelli:2008pk},
given the rich dynamical structure in the Standard Model (SM).  
In particular, stability or longevity of DM particles could be associated 
with exact or approximate quantum numbers, that might be in turn the results of 
exact gauge symmetries or accidental 
symmetries of underlying dark gauge groups, in analogy with the electron stability 
and the proton longevity (see Refs. \cite{Baek:2013qwa, KO:2016gxk}  
for discussions along this line of thoughts). 

The simplest candidate for the new dark gauge interaction arises from a hidden $U(1)$ 
gauge group which is kinetically mixed with the $U(1)$ sector in the SM. 
The study of a new $U(1)$ gauge group has a long history~\cite{Holdom:1985ag}.
The operation of the LHC has provided a unique opportunity to test 
various scenarios with new $U(1)$ interactions~\cite{Cohen:2017pzm}. Here we
are interested in a light DM charged under a dark $U(1)_d$ group with
the mass around the sub-GeV scale.  As this light DM is produced 
energetically at a collider, it radiates multiple collimated
$U(1)_d$ gauge bosons, i.e., dark photons, which then decay back into
SM particles, forming detectable leptonic or hadronic jets.  This is an
analogue to the phenomenon called the parton shower in the SM, 
especially the electroweak~(EW) shower~\cite{Chen:2016wkt}, if the dark photon 
has a small mass. The subject on the dark shower has been investigated  recently
in the  literature: both analytical and Monte Carlo methods were applied to the 
model, in which DM fermions interact with gauge fields only through a vector 
current~\cite{Buschmann:2015awa}; both  the vector  and the axial vector 
interactions were considered in~\cite{Zhang:2016sll}, where it was pointed 
out that whether left-handed and right-handed fermions have different 
interactions with gauge bosons could be determined by measuring the dark
shower patterns at the LHC.

In this paper we will further explore the relation between the dark shower
patterns and the chirality of the DM fermion, motivated by the close 
connection of the DM property under the gauge group to the mass origin 
in the dark sector. The dark photon mass can come from two 
types of mechanism, the Higgs mechanism and 
the St\"{u}eckelberg mechanism~\cite{Bell:2016uhg,Ruegg:2003ps}. 
The latter can be seen as  a limiting case of the former with the Higgs 
sector -- including the longitudinal gauge boson after symmetry breaking 
and the Higgs boson -- decoupling from a theory, such that the fermion 
involved in the former (the latter) prefers to be chiral-like (vector-like).  
The dark shower pattern is then mainly governed by  transversely polarized 
dark photons in the case of  the St\"{u}eckelberg mechanism,  but receives 
additional contributions from longitudinally polarized dark photons
and dark Higgs bosons in the case of  the Higgs mechanism.
Because transversely polarized dark
photons tend to be soft, while longitudinally
polarized dark photons and dark Higgs bosons do not, 
different shower patterns can be produced in the two scenarios.
Therefore, exploring the chiral behavior of the DM fermion through the 
dark shower patterns helps understand the origin of the dark photon mass.

The rest of the paper is organized as follows. In Sec.~II, we elaborate
the two mass generation mechanisms for dark photons  and 
how they are related to the chiral property of the DM fermion. Two simplified 
models, the Chiral Model and the Vector Model, are introduced for the
realization of mass generation. In Sec.~III, we explain the setting 
of the dark shower and the role of the splitting functions, mentioning 
some subtleties attributed to particle mass effects. The splitting 
functions with the DM fermions as the initial particles in the considered models
are then derived according to the formalism for the EW shower in 
Ref.~\cite{Chen:2016wkt}. In Sec.~IV,  we implement the dark shower with the
Monte Carlo program developed in Ref.~\cite{Chen:2016wkt}, 
examine several observables associated with the dark shower, and
highlight the different patterns between the two models. It will
be demonstrated that the DM jet energy profile, being  
almost independent of the DM energy, is an appropriate observable
for differentiating the DM fermion chirality. We intend to explore the 
properties of new dark $U(1)$ gauge boson showers possibly produced at LHC 
as an application of the results in~\cite{Chen:2016wkt}, and to lay 
out a correct framework for studying this topic. We emphasize that  
there has not been a complete treatment of the dark splitting functions and the
dark shower implementation in the literature. Section~V is the conclusion.  
Some examples on the calculation of the splitting functions are presented
in the Appendix.

\section{Models}

A peculiar observation about an abelian gauge theory is that a gauge boson
can obtain a mass without the Higgs mechanism,  while the theory
still remains gauge invariant and renormalizable.  The mechanism is
referred to as the St\"{u}eckelberg mechanism~\cite{Bell:2016uhg,Ruegg:2003ps}
which differs from the well-known Higgs mechanism in the number of degrees of freedom. 
The latter requires an additional scalar
field charged under the gauge group to induce symmetry breaking, after
which the Goldstone modes are ``eaten'' by the gauge fields to become the
longitudinal polarizations.  As the dark photon mass is
generated through the Higgs mechanism, there are effectively two more
degrees of freedom, the longitudinal polarization of the dark photon and 
the dark Higgs boson.  The St\"{u}eckelberg mechanism
is a limiting case of the Higgs mechanism, in which the vacuum expectation
value~(VEV) of the Higgs boson approaches to infinity, while the Higgs charge and
the Yukawa coupling approach to zero in the way that the gauge boson (fermion) mass,
proportional to the product of the Higgs charge (Yukawa coupling) and the VEV, 
remains fixed. The Higgs boson, with its mass being 
proportional to the product of the square root of the finite Higgs self-coupling and 
the VEV, then decouples. Hence,
if the dark photon obtains its mass through
the St\"{u}eckelberg mechanism, neither the Goldstone mode nor the dark 
Higgs boson will exist.

As stated in the introduction, the origin of the dark photon mass is closely 
related to the DM fermion property under the gauge group $U(1)_d$.  The 
argument goes as follows:
we first assume that the DM fermion is of the Dirac type and has some 
generic interactions with the dark photon. If the DM fermion is chiral-like,  
the left-handed fermion and the right-handed fermion can have different 
$U(1)_d$ charges, and  a bare mass term for the fermion is forbidden by the 
$U(1)_d$ symmetry. Similarly to the SM,  a dark Higgs field has to be introduced 
to give the fermion mass, which then gives the dark photon mass as well 
naturally. Thus the dark  photon mass is likely to be induced by the Higgs 
mechanism in this case. Alternatively, if the DM fermion is vector-like, 
the left-handed and right-handed fermions have the same charge under the
dark $U(1)_d$ group. It is then impossible for the fermion mass to come from 
the symmetry breaking of a Higgs sector under the same $U(1)_d$ group. 
It is also natural to assume that the dark photon mass is attributed to
the St\"{u}eckelberg mechanism without a Higgs sector.

We realize the above two scenarios with the simplified models below.
The Chiral Model for the implementation of the Higgs mechanism is defined as
\be
{\cal L}  &=&  -\frac{1}{4} F'_{\mu\nu} F'^{\mu\nu}  + \frac{\epsilon}{2} F'_{\mu\nu} F^{\mu\nu} 
+ | D_\mu \Phi' |^2 - \frac{\lambda_{\Phi'}}{4} \left( | \Phi '|^2 - \frac{v_{\Phi'}^2}{2} \right)^2   \n \\ 
& &+ \sum_{s=L/R}i \overline{\chi}_s  \slashed{D}   \chi_s -  \left( y_\chi \overline{\chi_{L}} 
\Phi' \chi_{R} + h.c. \right), \label{chi}
\ee
where the fields with primes represent the dark fields, $\epsilon$
describes the mixing strength between the dark and SM photons,
$D_{\mu } = \partial_\mu - i g' Q_{s}A'_\mu$ with $s=L/R$ for
the left-/right-handed DM fermion $\chi_s$, the Higgs charge $Q_{\Phi'}$
appearing in $D_\mu \Phi'$ is given by $Q_{\Phi'}=Q_L-Q_R$, and
$\lambda_{\Phi'}$ and $y_\chi$ denote the dark Higgs self-coupling
and the dark Yukawa coupling, respectively.
The scalar field can be parameterized as $\Phi'=\frac{1}{\sqrt{2}}(h'+i\phi')$.  
After dark gauge symmetry breaking,   $\Phi'$ acquires a VEV $v_{\Phi'}$ along the 
direction of $h'$: $h' \rightarrow h' + v_{\Phi'}$, and particles get 
their masses with the dark photon mass $m_{A'}=g'Q_{\Phi'}v_{\Phi'}$,  
the dark fermion mass $m_{\chi}=\frac{y_{\chi}v_{\Phi'}}{\sqrt{2}}$, 
and the dark Higgs mass $m_{h'}^2=\frac{\lambda_{\Phi'}v_{\Phi'}^2}{2}$.  
Here we have adopted the sign convention of the coupling, so that
$g'Q_{\Phi'}>0$ and $y_{\chi}>0$. 

It is easy to see that Eq.~(\ref{chi}) reduces to the Vector Model for
the St\"{u}eckelberg mechanism in the limits $v_{\Phi'}\to \infty$, 
$Q_{\Phi'}\to 0$, and $y_{\chi}\to 0$ with finite $m_{A'}$ and $m_{\chi}$,
\begin{equation}
{\cal L}  =  -\frac{1}{4} F'_{\mu\nu} F'^{\mu\nu}  
+ \frac{\epsilon}{2} F'_{\mu\nu} F^{\mu\nu} 
+ \frac{1}{2} m_{A'}^2 A'_\mu A'^\mu 
+ \sum_{s}\overline{\chi}_s \left( i \slashed{D}  - m_\chi \right) \chi_s,
\end{equation}
for which we have $Q_L=Q_R$. 

The two models are typical, and do not cover all the 
possibilities~\cite{Bell:2016uhg}. In the chiral case, other possibilities 
are highly constrained by the unitarity and gauge 
invariance~\cite{Kahlhoefer:2015bea}, such that a dark Higgs 
sector seems to be inevitable.
In the vector case, the dark photon mass is still allowed to arise from the 
Higgs mechanism, but
we would need to add additional degrees of freedom to the model. 
Because these possibilities do not modify the relation between the shower 
patterns and the DM fermion chirality essentially, we will ignore them here
without losing generality, and leave them to future works.

\section{Collinear Splitting functions and Dark Shower}

\subsection{Mass Effects}

When the masses of the DM  and the dark photon are much lower than 
the center-of-mass energy of a collider, their production rates are 
greatly enhanced in collinear regions of radiative corrections, 
leading to multiple dark particles collimated with the DM along a 
certain direction.
This dark shower is in analogy to the QCD and EW showers in the SM.
If the dark photon has a finite mixing with the SM photon, 
the produced dark photons may decay into SM particles, resulting in 
signatures of lepton jets~\cite{Buschmann:2015awa} or light-hadron jets.

\begin{figure}[t]
\label{darkshower}
\centering
\includegraphics[width=10cm]{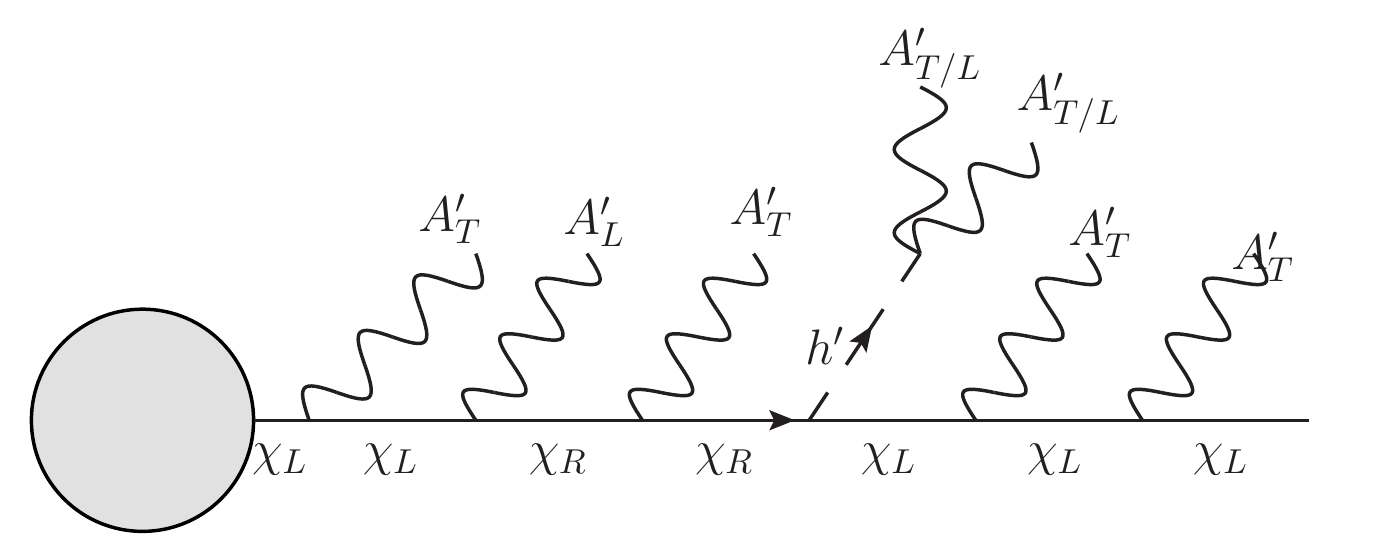}
\caption{Illustration of dark shower in the Chiral Model.}
\end{figure}

The evolution of the dark shower initiated by the
mother particle $A$ through the radiation $A\rightarrow B+C$
is controlled by the Sudakov form factor
\be\label{eq:sudakov}
\Delta_A(t)= \text{exp}\left[-\sum_{B,C}\int_{t_0}^t dt' 
\int dz \frac{d{\mathcal P}_{A\rightarrow B+C}}{dzdt'}\right],
\ee
\noindent 
which sums all possible collinear splitting functions
$\frac{d{\mathcal P}_{A\rightarrow B+C}}{dzdt'}$.  
The variable $z=\frac{E_C}{E_A}$ is the energy fraction of the
particle $C$ to the particle $A$. 
The evolution variable is usually taken as  $t=\log
(k_T^2)$ or $t=\log (q^2)$ with $k_T$ being the transverse momentum
of a final state particle and $q^2$ being the virtuality of $A$. The lower bound
$t_0=\log (m_{\text{cutoff}}^2)$ corresponds to the infrared cutoff scale
$m_{\text{cutoff}}={\rm max}(m_{A'}, m_{\chi},m_{h'})$.  
As seen below,  the mass terms in the splitting functions play the role of
an infrared cutoff, so that the choice of $t_0$ is largely irrelevant as long
as it is not higher than  the mass scale of the theory.

The dark shower in a massive $U(1)$ theory bears many
similarities to the EW shower. In Ref.~\cite{Chen:2016wkt}, all the
$1\rightarrow 2$ EW splitting functions were derived, including the
broken splitting functions that are proportional to the VEV of the Higgs field,
or equivalently, particle masses. The splitting function 
$\frac{d{\mathcal P}_{A\rightarrow B+C}}{dzdk_T^2}$ can be expanded 
in powers of $\frac{m^2}{k_T^2}$~\cite{Chen:2016wkt} in a model 
with symmetry breaking,  
\be
\text{leading power}: \   \   \   \frac{d{\mathcal P}_{A\rightarrow B+C}}{dzdk_T^2} 
&\propto& 
\frac{k_T^2}{\tilde{k}_T^4}, \label{eq:splittings} \\
\text{next-to-leading power: }\  \  \   \frac{d{\mathcal P}_{A\rightarrow
B+C}}{dzdk_T^2} &\propto& 
\frac{m^2}{\tilde{k}_T^4}, \label{eq:broken}
\ee
\noindent  
where the mass parameter $m$ 
depends on the specific splitting process. 
The denominator $\tilde{k}_T^4=(\tilde{k}_T^2)^2$ 
is written in terms of  
\be
\tilde{k}_T^2= k_T^2+\bar{z}m_B^2+z m_C^2-z\zb m_A^2, 
\ee
with $\zb=1-z$.
The splitting functions at the leading power, being mass independent, 
correspond to those in the unbroken theory. The splitting functions 
from the next-to-leading-power corrections are more 
enhanced at low $k_T$ relative to the unbroken splittings, and called 
the ``ultra-collinear" 
splittings~\cite{Chen:2016wkt}.  The origin of the ultra-collinear 
splittings can be interpreted as the VEV insertions into either
particle propagators or splitting vertices~\cite{Chen:2016wkt}.

Compared to the splitting functions for massless particles, 
we have replaced $\frac{1}{k_T^2}$ by $\frac{k_T^2}{\tilde{k}_T^4}$ effectively, 
such that the mass terms in $\tilde{k}_T^4$ play the 
role of an infrared regulator. The evolution of the parton shower will 
shut off automatically, when it approaches to the infrared scale. Note 
that the infrared regularization in the QCD shower is implemented with
a sharp cutoff, below which the hadronization takes place. The mass 
effects are included in {\tt Pythia}~\cite{Sjostrand:2014zea} currently
by adding an extra term to the splitting function~\cite{Buschmann:2015awa},  
\be
\frac{1}{k_T^2} \rightarrow \frac{1}{k_T^2}+\frac{m^2}{k_T^4},
\ee
\noindent equivalent to the Taylor expansion of 
$\frac{k_T^2}{\tilde{k}_T^4}$ around  $\frac{1}{k_T^2}$ to the order of 
$\frac{m^2}{k_T^2}$.

The separation of the unbroken and broken pieces is best illustrated
in the splittings containing longitudinal vector bosons. Naively, the 
splitting function for $\chi\rightarrow \chi A'_L$ can be obtained through 
the Goldstone equivalence theorem, whose contribution, however, accounts 
only for the unbroken piece. It has been proposed to take into account  
the symmetry breaking effects by imposing the Goldstone 
equivalence gauge (GEG)~\cite{Chen:2016wkt}.  To explain what this new 
gauge does, we write the longitudinal polarization vector as 
\be
\epsilon^{\mu}_L=\frac{k^{\mu}}{m_{A'}}-\frac{m_{A'}}{n\cdot k}n^{\mu},
\ee
with the  momentum of the vector boson  $k^{\mu}=(k^0, \vec{k})$ and 
the direction of GEG being defined by a null vector $n^{\mu}=(1, -\hat{k})$ 
with $\hat{k} \cdot \hat{k} = 1$.
The term $\frac{k^{\mu}}{m_{A'}}$ is the one that gives rise to the 
aforementioned contribution of the Goldstone equivalence. It  
induces a bad high-energy behavior and large interference among diagrams, 
complicating many calculations, such as those of the collinear splitting 
functions. Working in the GEG along $n^{\mu}$ renders
this term, which violates the gauge condition 
because of $n\cdot k\neq 0$, not contribute to physical polarizations. 
Instead, it manifests itself as a Goldstone mode.  The 
remnant term $-\frac{m_{A'}}{n\cdot k}n^{\mu}$ survives, since 
$n\cdot n=0$, namely, the gauge condition is satisfied. 
The amplitudes involving longitudinal vector bosons
are then evaluated by summing diagrams for both the Goldstone and gauge
components in GEG. These two components bear different physical 
significance to the splitting functions: the former,  
that flips the fermion helicity, contributes to  
splittings at leading power of $\frac{m^2}{k_T^2}$; while the latter, 
that does not flip the fermion helicity, 
contributes at next-to-leading 
power, i.e., to the ultra-collinear splittings as seen in the next 
subsection.  Besides, the fermion mass
also contributes to the ultra-collinear splittings in a similar way.

\subsection{Splitting Functions}
\label{subsec:splitting}

The splitting functions in the Chiral Model and Vector Model are 
described by the same set of parameters 
$\alpha'=g^{\prime 2}/(4\pi)$, $m_{A'}$, $m_{\chi}$, $m_{h'}$, 
as well as $Q_{L}$ and $Q_{R}$,
in terms of which all other parameters $Q_{\Phi'}$, $y_{\chi}$, 
and $\lambda_{\Phi'}$ can be expressed.
We focus on the splittings with $\chi$ being the only initial state 
in the present work. The leading power splittings are given by 
\be
\frac{d{\mathcal P}}{dzdk_T^2}(\chi_s\rightarrow \chi_s+A'_T)
&=& \frac{\alpha'}{2\pi}Q_s^{2}\frac{1+\zb^2}{z} \frac{k_T^2}{\tilde{k}_T^4},  \\
\frac{d{\mathcal P}}{dzdk_T^2}(\chi_s\rightarrow \chi_{-s}+A'_L
)&=&\frac{\alpha'}{2\pi}\frac{m_{\chi}^2}{m_{A'}^2}Q_{\Phi'}^{2}
\frac{z}{2}\frac{k_T^2}{\tilde{k}_T^4},\label{eq:split_unbroken} \\
\frac{d{\mathcal P}}{dzdk_T^2}(\chi_s\rightarrow \chi_{-s}+h')
&=&\frac{\alpha'}{2\pi}\frac{m_{\chi}^2}{m_{A'}^2}Q_{\Phi'}^{2}\frac{z}{2}
\frac{k_T^2}{\tilde{k}_T^4},
\ee
where $s$ denotes both the helicity $\pm1/2$ in $\chi_s$ and the chirality 
$L/R$ in $Q_s$. The helicity and the chirality become identical in the high 
energy limit with $s=\mp\frac{1}{2}$ corresponding to $s=L/R$ for particles 
(as opposed to antiparticles). Here we use left-handed/right-handed to 
label the helicity and the chirality interchangeably. It is found from the 
above splitting functions that the radiation of transversely polarized dark 
photons exhibits a soft enhancement at small $z$, and that the radiations 
of longitudinally polarized dark photons and dark Higgs bosons diminish at 
leading power in the Vector Model due to $Q_{\Phi'}=0$. These are the major 
features which cause the different dark shower pattens in the Chiral and Vector
Models.

We have the next-to-leading-power splitting functions
\be
\frac{d{\mathcal P}}{dzdk_T^2}(\chi_s\rightarrow \chi_{-s}+A'_T)
&=& \frac{\alpha'}{2\pi}z(Q_s-Q_{-s}\zb)^2 \frac{m_{\chi}^2}{\tilde{k}_T^4},  \\
\frac{d{\mathcal P}}{dzdk_T^2}(\chi_s\rightarrow \chi_{s}+A'_L)
&=&\frac{\alpha'}{2\pi}\frac{1}{2z} \left(2Q_s\zb+(-1)^{s+\frac{1}{2}}
\frac{z^2m_{\chi}^2}{m_{A'}^2}Q_{\Phi'}\right)^2 \frac{m_{A'}^2}{\tilde{k} _T^4}, \label{eq:split_broken}\\
\frac{d{\mathcal P}}{dzdk_T^2}(\chi_s\rightarrow \chi_{s}+h')
&=&\frac{\alpha'}{2\pi}Q_{\Phi'}^{2}\frac{z(1+\zb)^2}{2}
\frac{m_{\chi}^2}{m_{A'}^2}\frac{m_{\chi}^2}{\tilde{k}_T^4}.
\ee
At this subleading level, longitudinally polarized dark photons contribute 
in the Vector Model, but dark Higgs boson still do not.  As shown in 
the next section, the next-to-leading-power effects on the dark shower patterns 
are less important.

In the above derivation with only the dark radiation, we have assumed that
the mass eigenstate of the massive dark photon is what appears in  the
Lagrangian. Strictly speaking, we need to perform the field redefinition 
and diagonalize the mass matrix to find the real mass eigenstates first.
After the diagonalization, the real massless eigenstate 
does not interact with the DM fermion directly, and
the massive dark photon can be also radiated by a SM fermion, such as a 
colliding parton, whose effect is, however, suppressed by the mixing 
parameter $\epsilon$. Besides, the $1\rightarrow 2$ 
splitting amplitudes mainly collect collinear contributions, and it has 
been known that different collinear sub-processes do not affect each 
other significantly. Including a $U(1)_Y$ gauge group from the SM side, we get 
an additional interaction between the DM fermion and the $Z$ boson.
This interaction does not induce new collinear splittings, because the
$Z$ boson mass is much larger than the mass scale considered here.

Compared with Ref.~\cite{Zhang:2016sll} and the setting in {\tt Pythia},  
our formulae have several important differences:
\begin{itemize}  
\item  In the splittings, we treat the fermion helicities separately. 
This is necessary, because it is not guaranteed that the initial particle 
in the shower is unpolarized.  Moreover, the fermion flips its helicity 
in some splittings, leading to nontrivial interplay between different 
helicities, which cannot be captured by naively 
taking an average of the initial helicities in the splittings. 
Especially, we find that even though the DM is unpolarized
initially, it can obtain a certain polarization after showering in our 
setup\footnote{As an example let us take the benchmark point A for the Chiral Model 
in the numerical  analysis below. Starting from unpolarized DM fermions, 
we get roughly 70\% left-handed  DM fermions and 30\% 
right-handed DM fermions in the final states.}.
\item We incorporate the dark Higgs boson contribution in the splitting functions,  
since it arises 
naturally along with the Goldstone mode in the Chiral Model.  
\item Our splitting function for $\chi_s\rightarrow \chi_{-s} + A'_L$ contains an 
additional factor $z/2$ relative to the result in 
Ref.~\cite{Zhang:2016sll}, which arises from the choice of 
the wave function for the initial state fermion in the evaluation 
of the splitting functions. We point out that in order for 
proper factorization of the collinear splitting functions from hard 
processes, the ``on-shell'' wave function  is required regardless 
of the kinematics, as elaborated further in Appendix~\ref{sec:example}.  
\item We have one more set of splitting functions (scaling as $\frac{m^2}{k_T^4}$) 
attributed to the symmetry breaking, which are more enhanced in the small $k_T$ region 
than the leading-power  splitting functions. 
\end{itemize}

\section{Implementation of Dark Showering}
\label{sec:imp}

We implement the dark shower with the derived splitting functions using 
the EW shower program from Ref.~\cite{Chen:2016wkt}, and compare its 
patterns in the Chiral Model and Vector Model at several benchmark points. 
For the same couplings and masses, the difference between the two models 
is characterized by the charge ratio $Q_{L}/Q_{R}$. 
Following Ref.~\cite{Zhang:2016sll} for  an immediate comparison,  
we choose $(Q_V,Q_A)=(1,1)$ for the Chiral  Model and $(Q_V,Q_A)=(1,0)$ for the Vector
Model, where $Q_V=\frac{Q_{L}+Q_R}{2}$ and $Q_A=\frac{Q_L-Q_R}{2}$.
Except for the dark Higgs mass $m_{h'}$,  the other
parameters of the models are also the same as in Ref.~\cite{Zhang:2016sll}.
Three benchmark points A, B and C are selected as 
\be
\text{point A:} \  \  && \alpha'=0.3 \  \  \  \   m_{\chi}=0.7 \ \text{GeV}  
\  \ m_{A'}=0.4 \ \text{GeV} \  \  m_{h'} = 1.0 \  \text{GeV},  \n \\
\text{point B:} \  \  && \alpha'=0.15 \ \   \    m_{\chi}=1.0 
\  \text{GeV}  \  \ m_{A'}=0.4 \  \text{GeV} \  \  m_{h'} = 1.0 \ \text{GeV}, \n \\
\text{point C:} \  \   && \alpha'=0.075 \ \   m_{\chi}=1.4 \  \text{GeV}  
\  \ m_{A'}=0.4 \  \text{GeV} \  \  m_{h'} = 1.4 \ \text{GeV}, \n
\ee
in which the DM fermion and the dark photon with the masses 
of around sub-GeV are relatively light, and the Yukawa coupling is as large 
as possible, i.e., near the perturbative limit 
$\alpha'\frac{m_{\chi}^2}{m_{A'}^2}\lesssim1$. It has been shown~\cite{Zhang:2016sll}
that the study with the above model parameters is relevant for the LHC search.

We simulate the hard process of DM fermion pair production 
at the LHC with the center-of-mass energy $\sqrt{s}=14$~TeV through 
the effective operator $(\bar{q} \gamma^\mu q) (\bar{\chi} \gamma_\mu \chi)$,
requiring an associated jet to have a transverse momentum 
$p_T>200$ GeV.  After the dark shower, dark Higgs bosons in the 
final state are assumed to exclusively decay into pairs of dark photons, which
subsequently form electron pairs, muon pairs and pion pairs. 
For our choice of the dark photon mass, the decay branching ratios 
are set to $\text{Br}(A' \to e e)=\text{Br}(A' \to \mu\mu)= 0.45$ and 
$\text{Br}(A' \to \pi \pi) =0.1$, respectively~\cite{Alekhin:2015byh}. 
For simplicity, we also assume that the produced dark photons mostly decay 
into SM particles inside a collider. It then demands a large enough kinetic mixing 
$\epsilon \gtrsim 8.2 \times 10^{-6}$, so that $A'$ decays within
a length of $\sim \mathcal{O}(1)$ mm according to the total decay width 
$\Gamma_{A'} \sim \alpha_{\text{em}} \epsilon^2 M_{A'}$. On the other hand, 
the mixing effect should be small enough for justifying the neglect of the 
initial state dark radiation as noted before. The subtle cases, in which
the dark photons partially decay into SM particles, and the initial state
dark radiation contributes, will be studied elsewhere.

For the shower patterns,  we consider three observables: 
($i$) the scalar sum of transverse momenta $p_T$ over all produced dark photons, 
\be
H_T=\sum_{i=A'}|p_{T_i}|,\label{ht}
\ee
($ii$) the number $n_{A'}$ of dark photons per event, and ($iii$)  a jet substructure
called the energy profile of the DM jet.
Because the dark photons are highly boosted, $H_T$ gives the same result 
as the scalar sum of $p_T$ over all leptons and hadrons
from the dark photon decays. Though the distribution in $n_{A'}$
reflects the nature of the dark sector, strictly speaking, 
the photon number is not an infrared safe observable in the high energy limit. 
Equation~(\ref{eq:sudakov}) implies that the small $k_T$ region in
the splitting is favored, namely, the emitted particles tend to form a
jet along the direction of the DM fermion momentum. It has been known that 
jet substructures serve as a powerful tool to explore properties of
parent particles which lead jets. For example, it was proposed 
in~\cite{Kitadono:2014hna} to differentiate the helicity of an energetic 
top quark by means of its jet energy profile. It will be demonstrated that 
the Chiral Model and the Vector Model are distinguishable in the $H_T$ 
and $n_{A'}$ distributions, as well as in the jet energy profile.

\begin{figure}
\begin{subfigure}{0.49\textwidth}
\centering
\includegraphics[width=\textwidth]{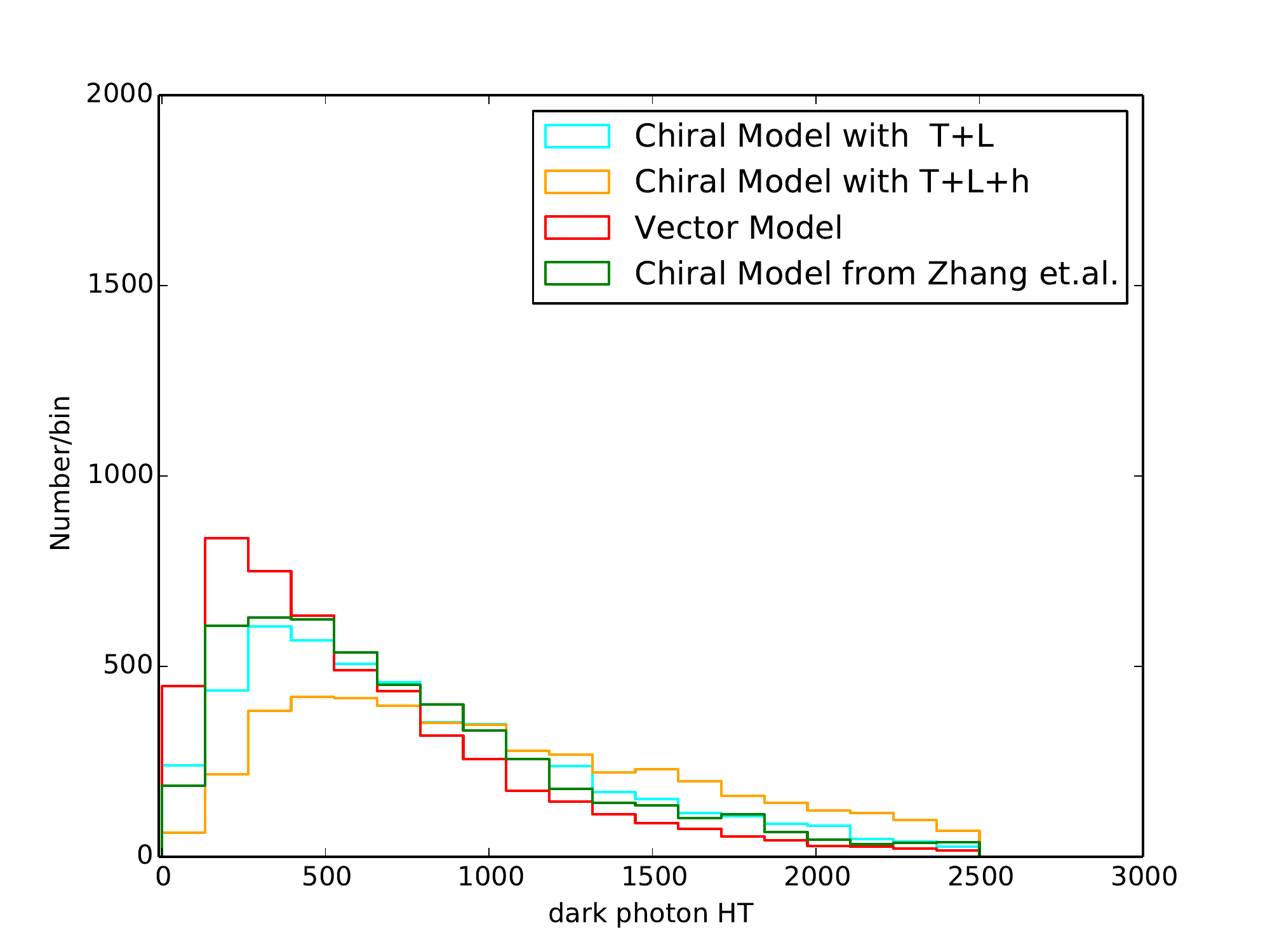}  
\caption{}
\end{subfigure}
\begin{subfigure}{0.49\textwidth}
\centering
\includegraphics[width=\textwidth]{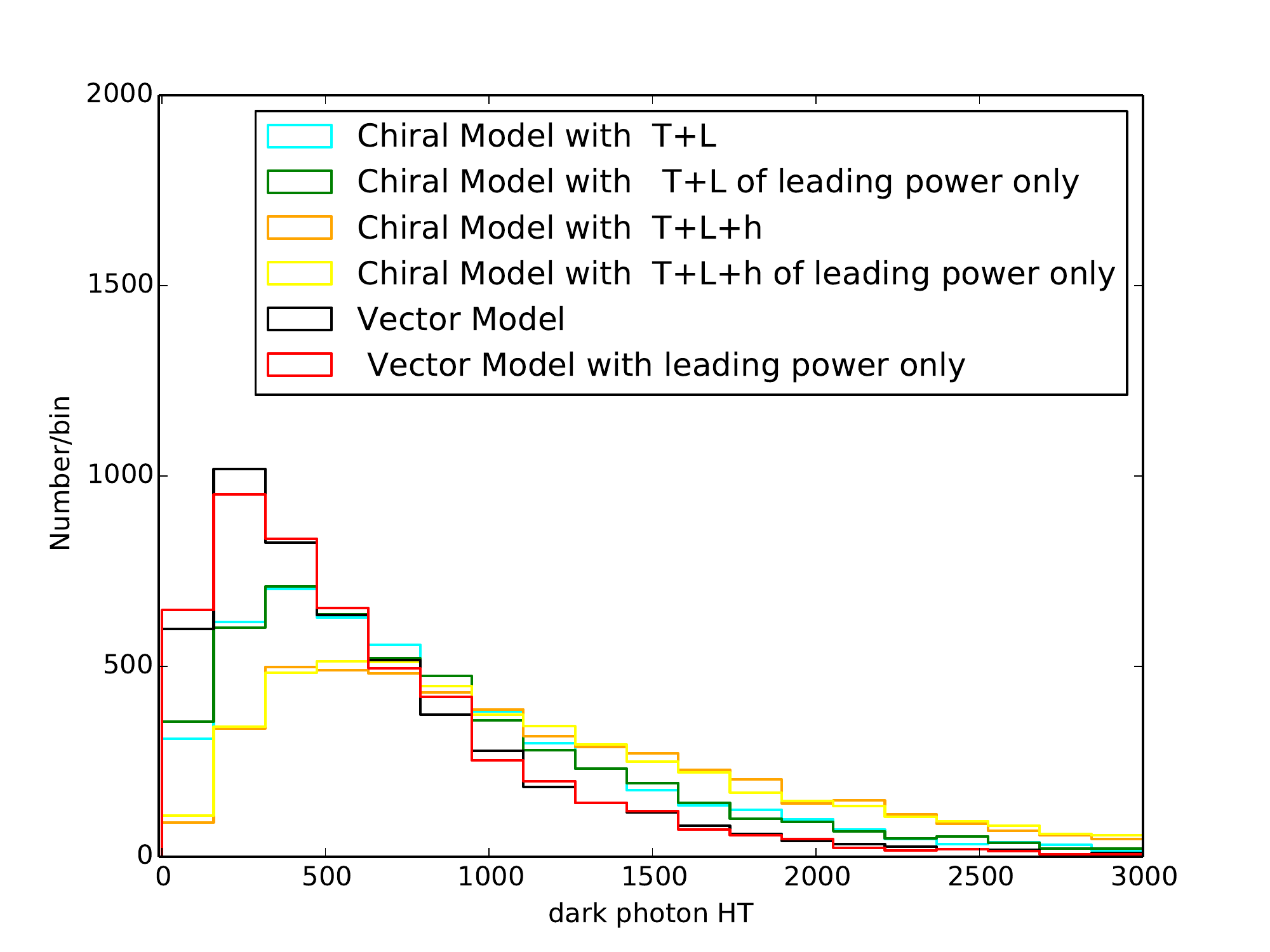}
\caption{}
\end{subfigure}

\begin{subfigure}{0.49\textwidth}
\centering
\includegraphics[width=\textwidth]{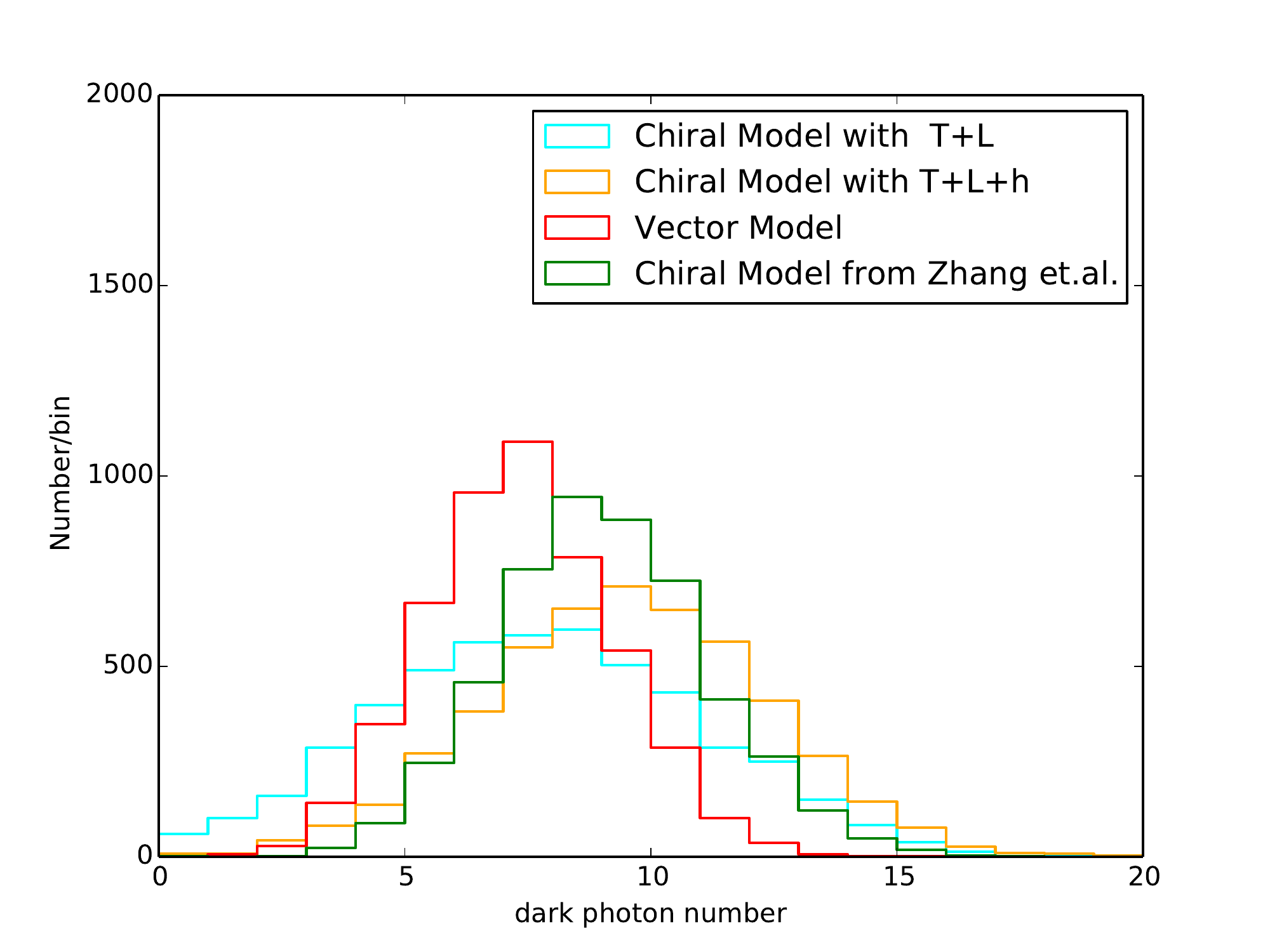}  
\caption{}
\end{subfigure}
\begin{subfigure}{0.49\textwidth}
\centering
\includegraphics[width=\textwidth]{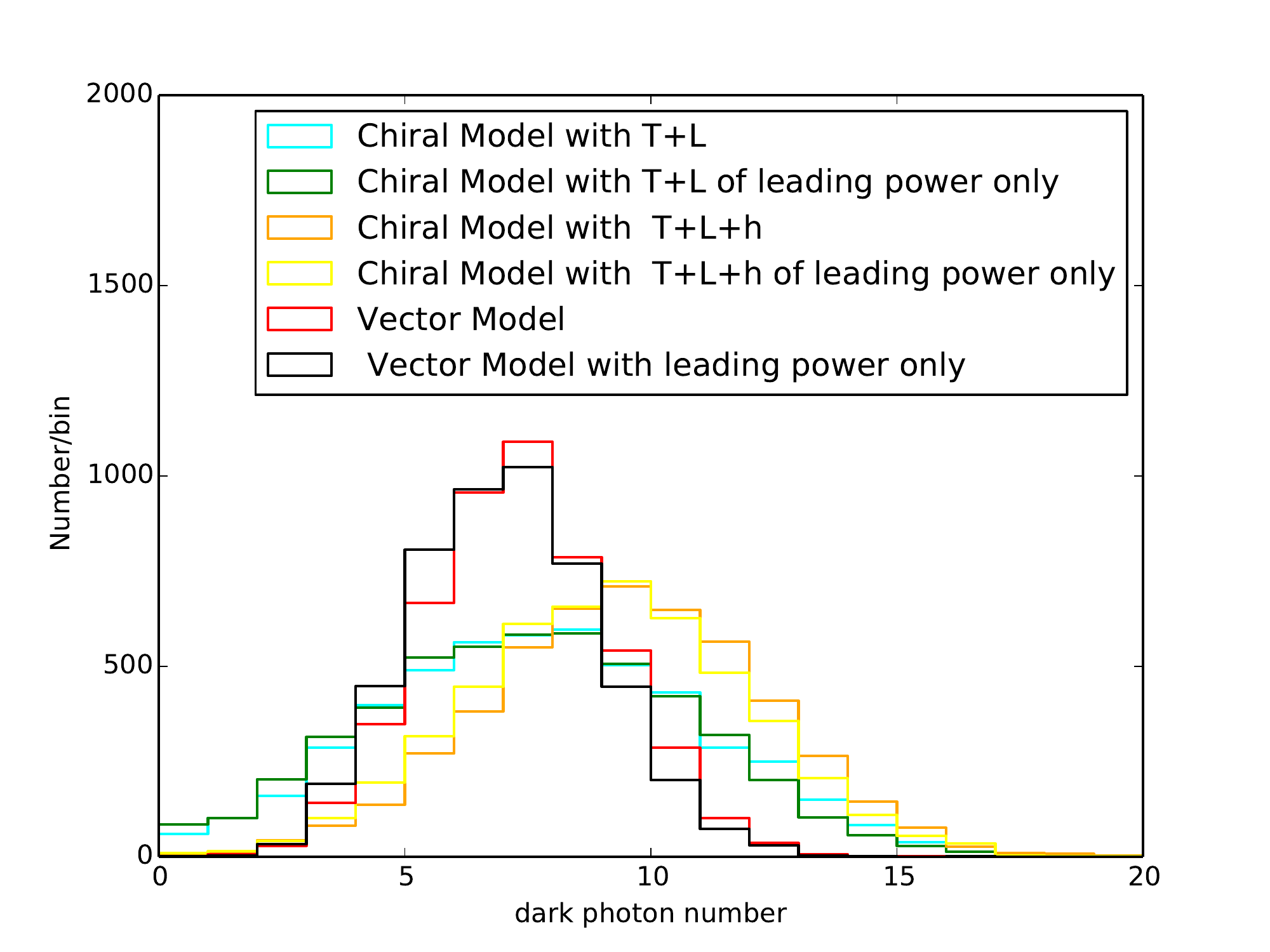}
\caption{}
\end{subfigure}
\caption{\label{fig:pointA} 
Dark shower with Point A: $\alpha'=0.3$, $m_{\chi}=0.7$ GeV, 
$m_{A'}=0.4$ GeV, and $m_{h'}=1.0$ GeV.   ``T", ``L", ``h" represent the 
types of splitting functions: ``T" for $\chi_s\rightarrow \chi_{s'} A'_T$; 
``L" for $\chi_s\rightarrow \chi_{s'} A'_L$; ``h" for 
$\chi_s\rightarrow \chi_{s'} h'$.  ``Leading power" denotes the 
splitting functions with leading power contributions, i.e.  those scaling
as $\frac{d{\mathcal P}}{dzdk_T^2}\sim \frac{1}{k_T^2}$.  
``Zhang et. al." labels the splitting functions from 
\cite{Zhang:2016sll} by Zhang et. al.  }
\end{figure}

\begin{figure}
\begin{subfigure}{0.49\textwidth}
\centering
\includegraphics[width=\textwidth]{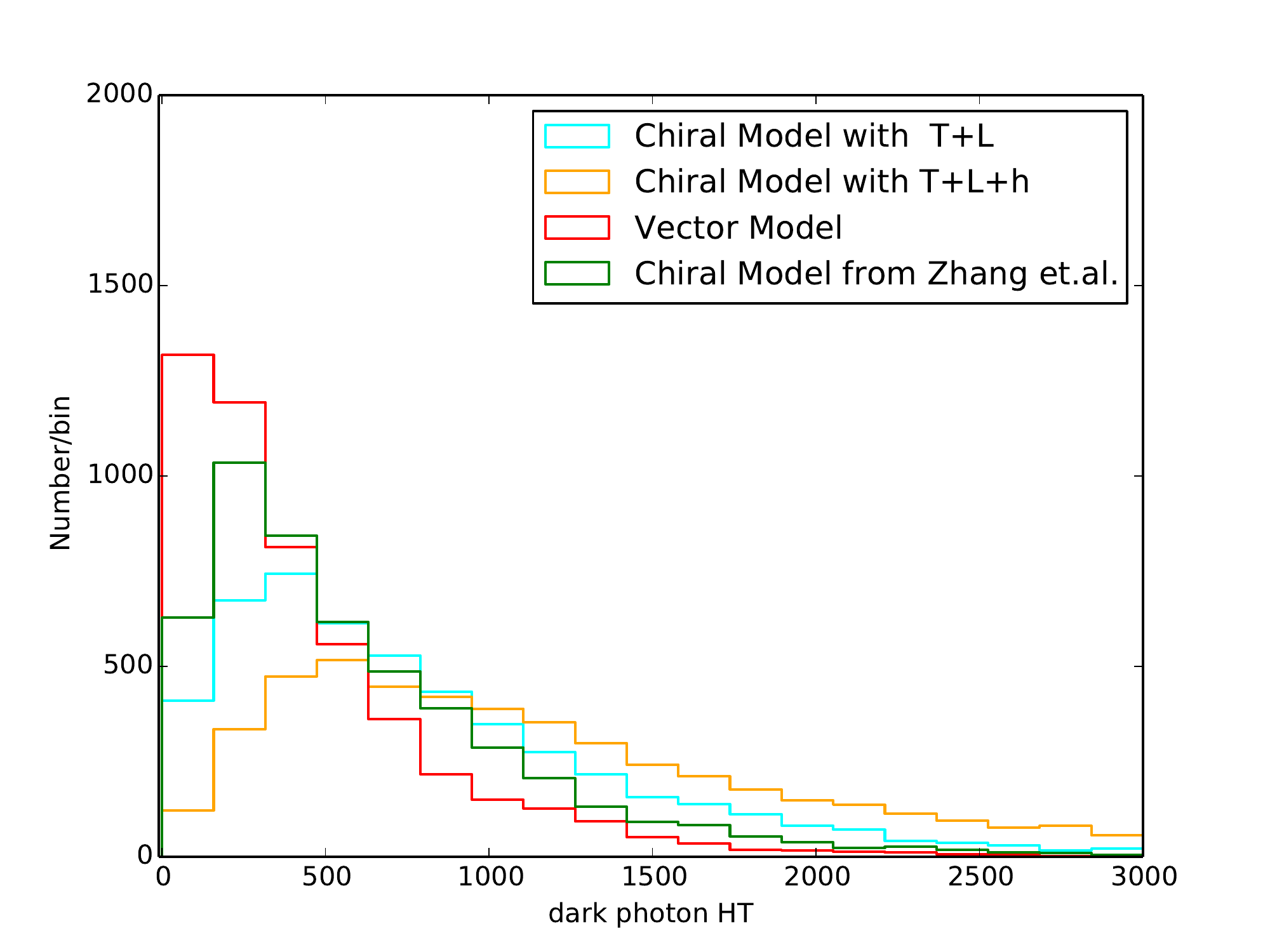}  
\caption{}
\end{subfigure}
\begin{subfigure}{0.49\textwidth}
\centering
\includegraphics[width=\textwidth]{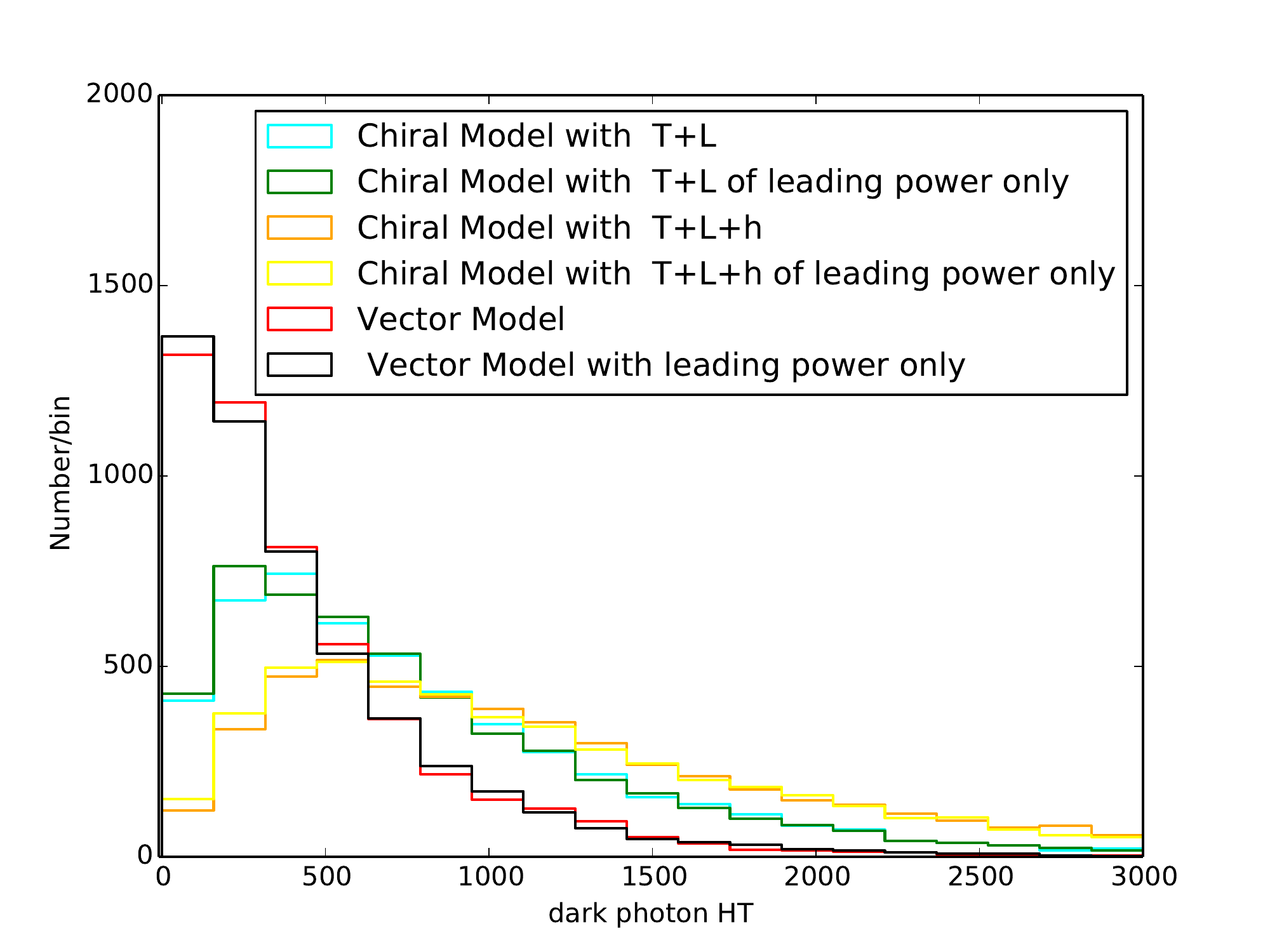}
\caption{}
\end{subfigure}

\begin{subfigure}{0.49\textwidth}
\centering
\includegraphics[width=\textwidth]{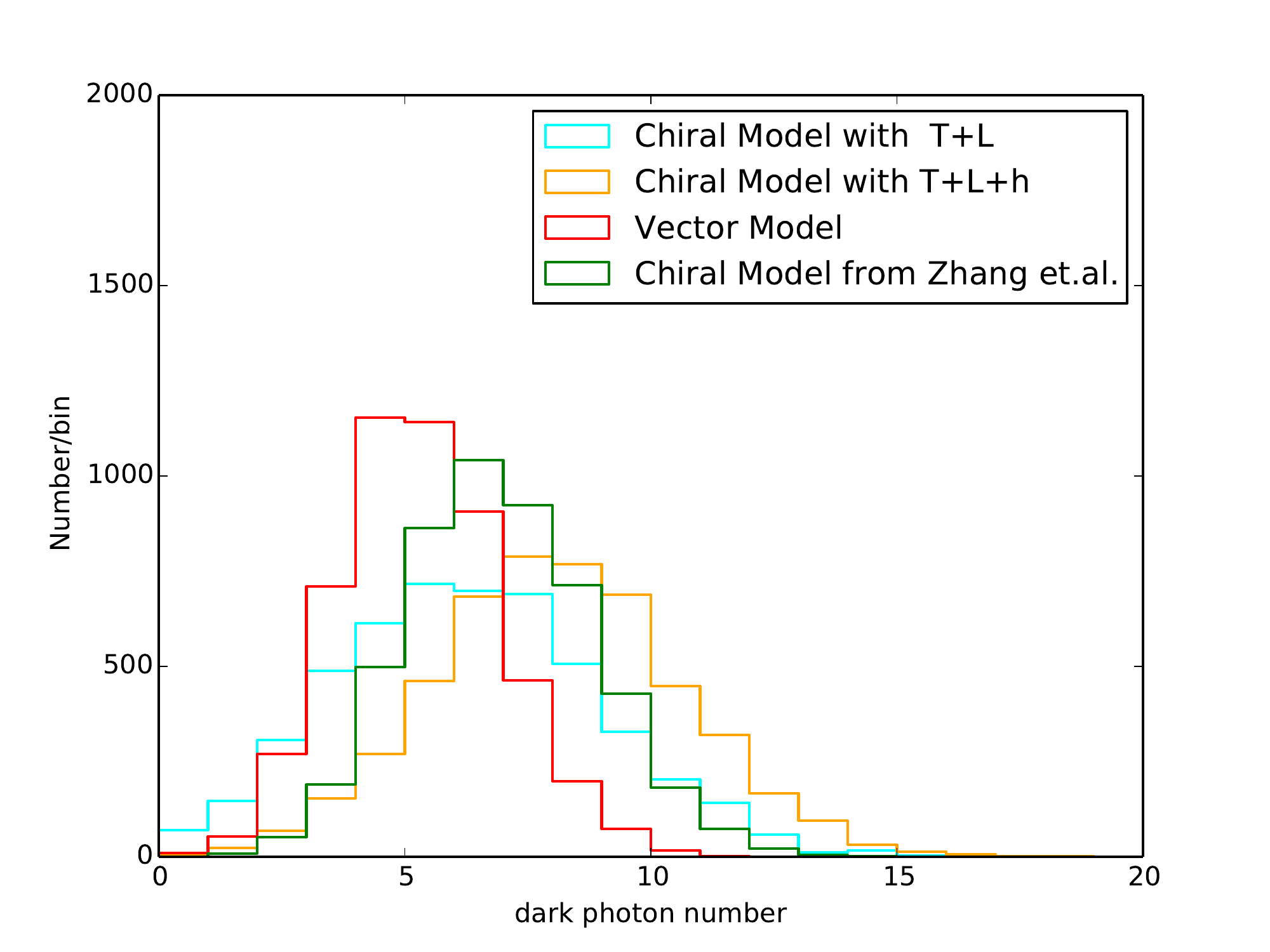}  
\caption{}
\end{subfigure}
\begin{subfigure}{0.49\textwidth}
\centering
\includegraphics[width=\textwidth]{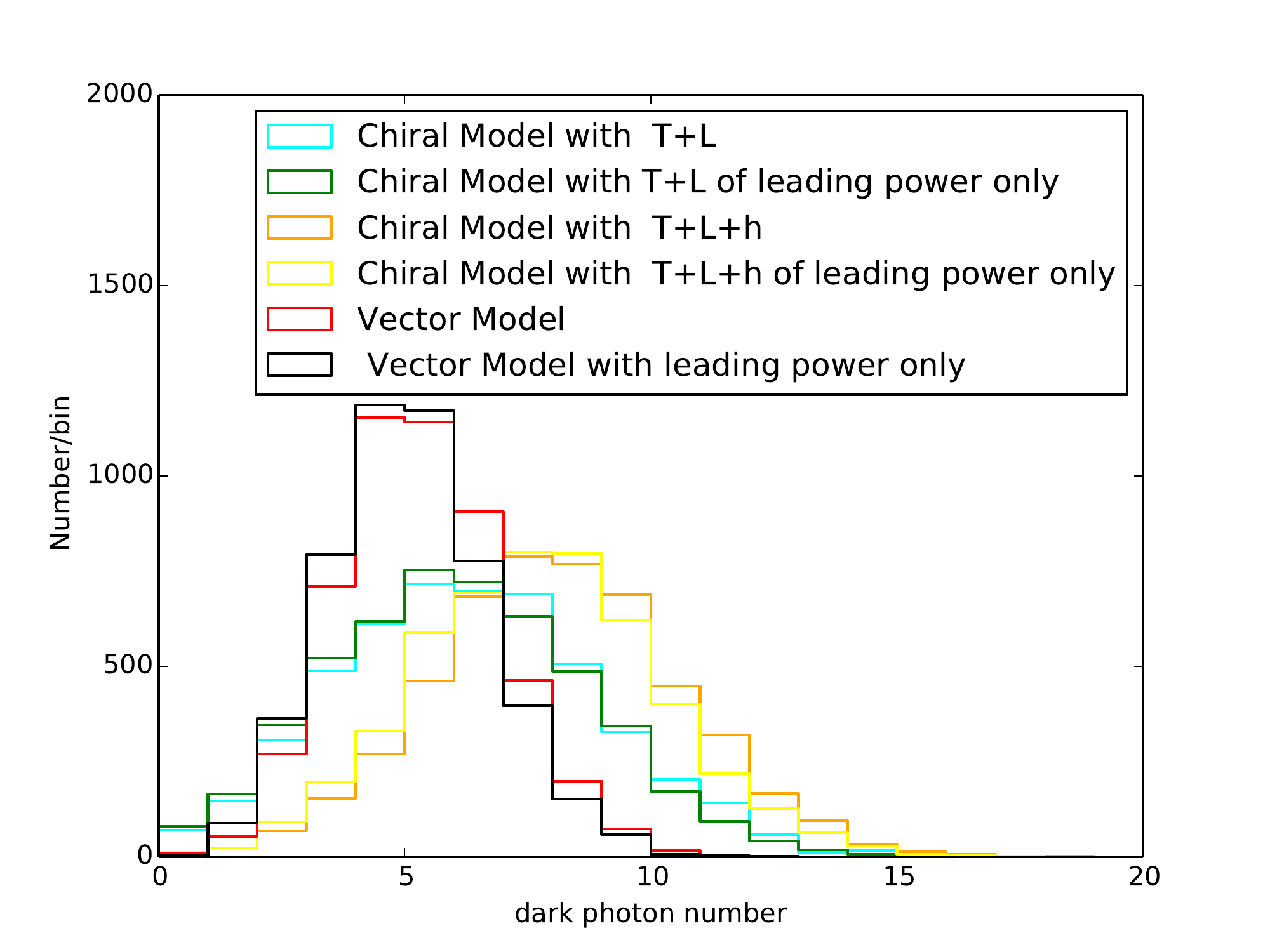}
\caption{}
\end{subfigure}
\caption{Same as Fig.~\ref{fig:pointA} but with Point B: $\alpha'=0.15$, 
$m_{\chi}=1.0$ GeV, $m_{A'}=0.4$ GeV, and $m_{h'}=1.0$ GeV. }
\label{fig:pointB}
\end{figure}

\begin{figure}
\begin{subfigure}{0.49\textwidth}
\includegraphics[width=\textwidth]{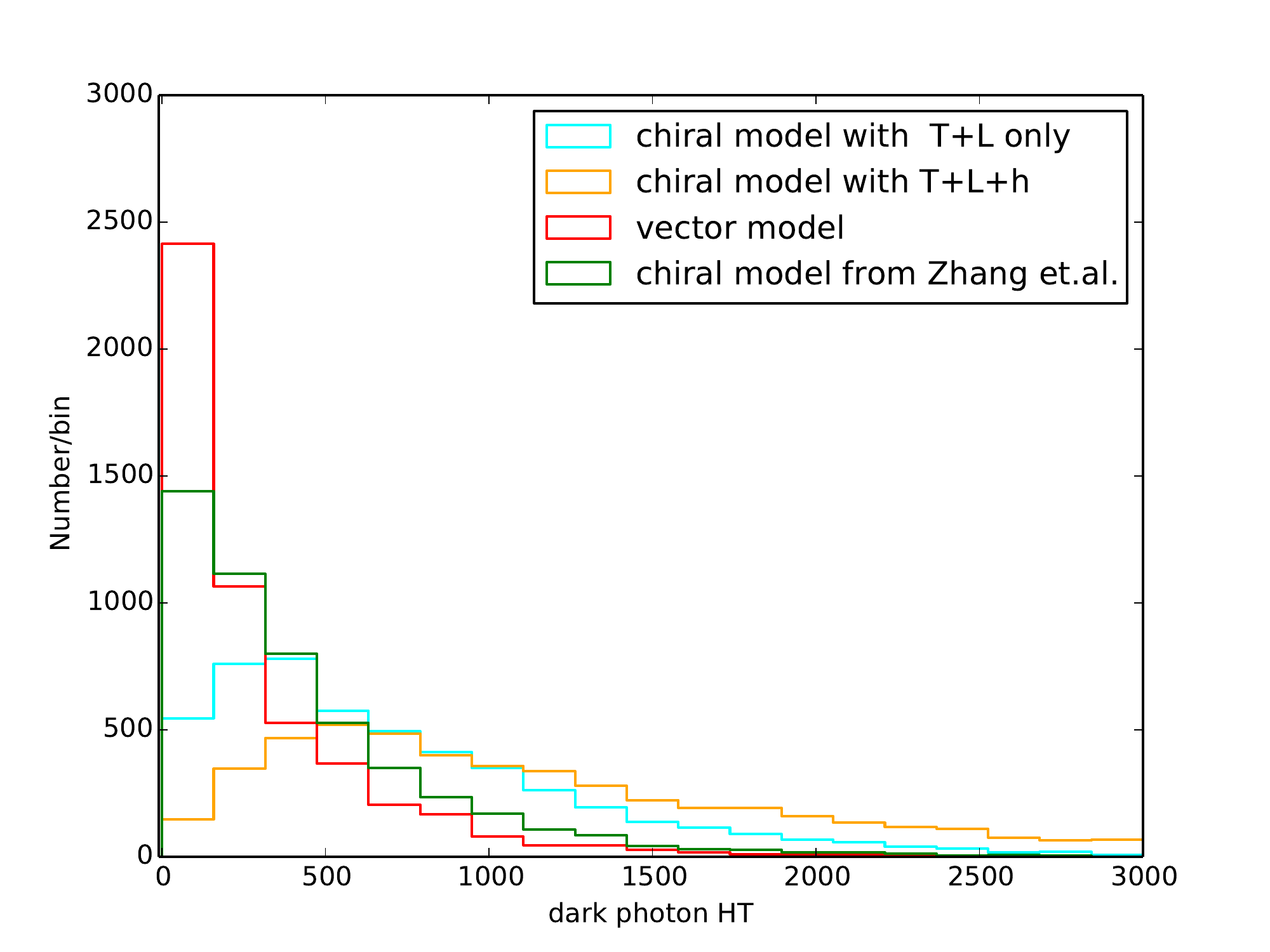}  
\caption{}
\end{subfigure}
\begin{subfigure}{0.49\textwidth}
\includegraphics[width=\textwidth]{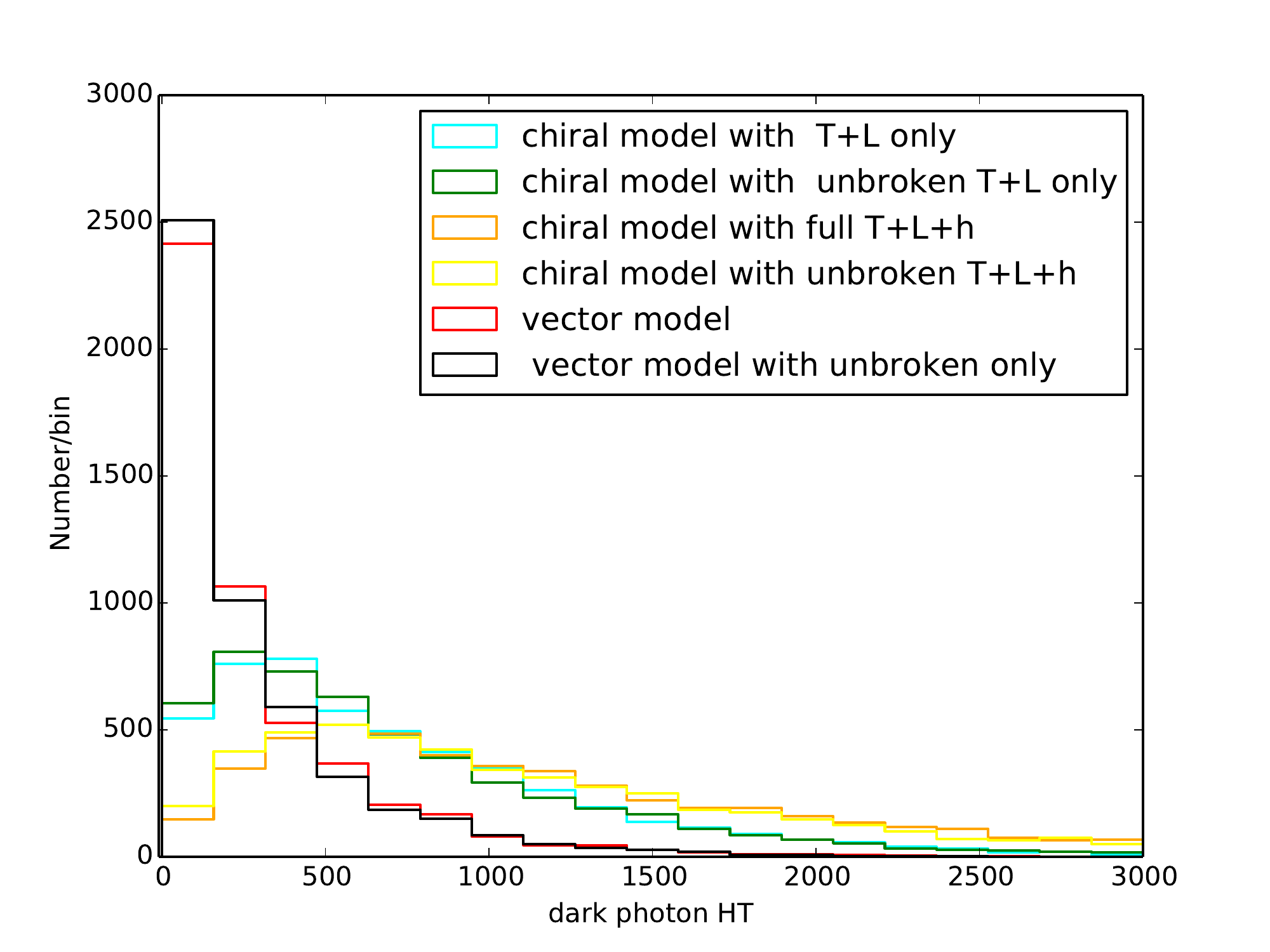}
\caption{}
\end{subfigure}
\begin{subfigure}{0.49\textwidth}
\centering
\includegraphics[width=\textwidth]{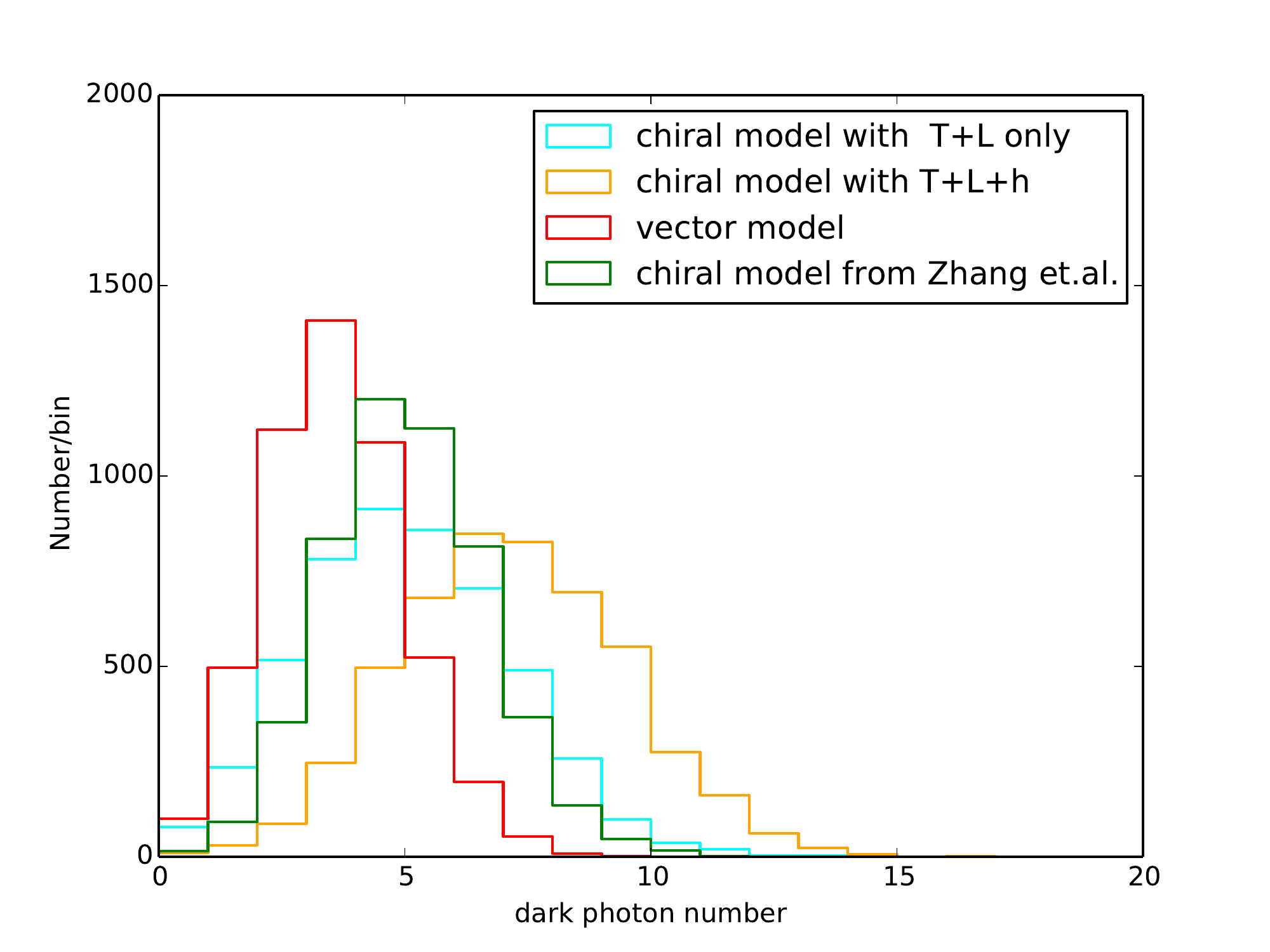}  
\caption{}
\end{subfigure}
\begin{subfigure}{0.49\textwidth}
\centering
\includegraphics[width=\textwidth]{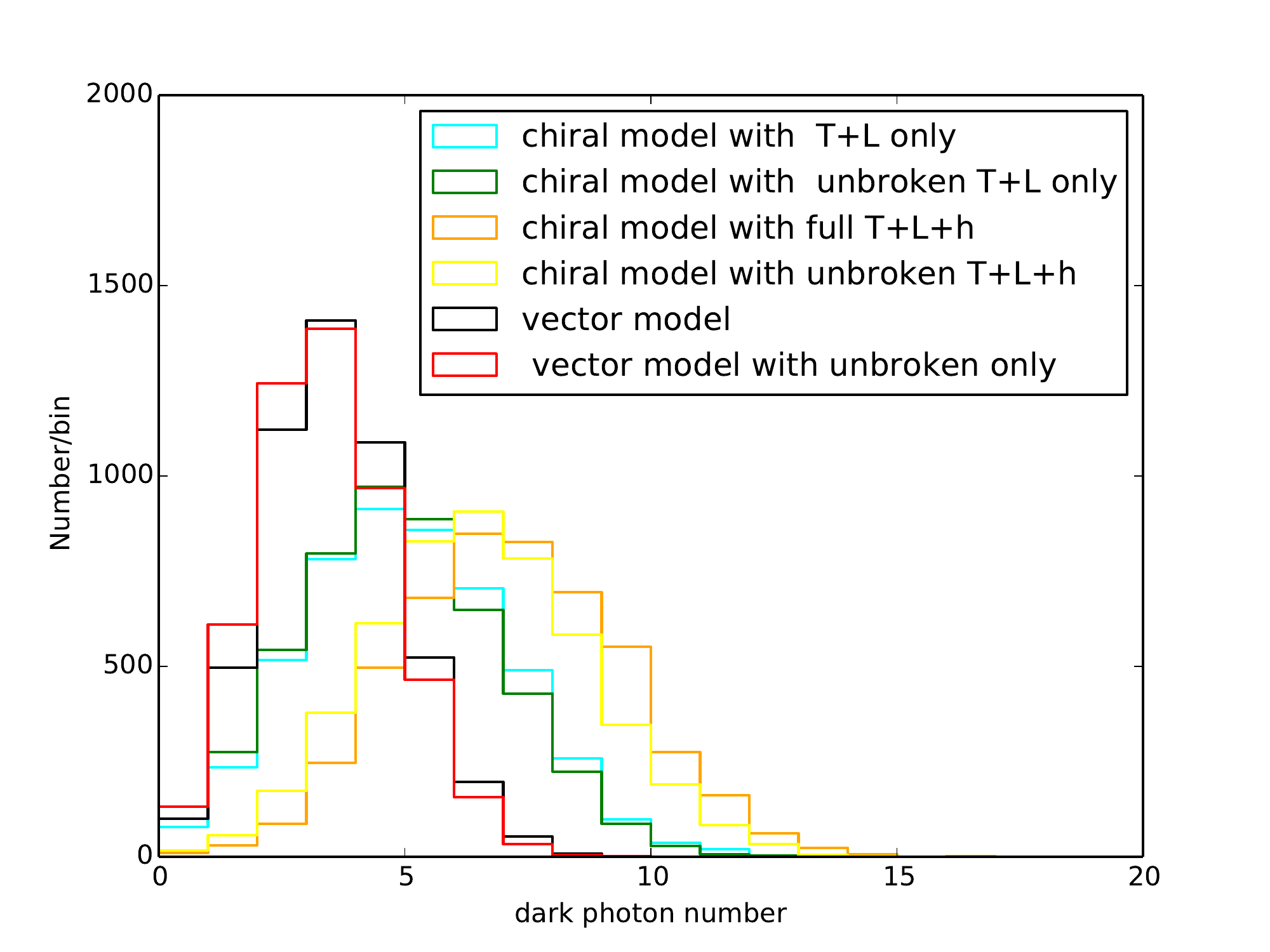}
\caption{}
\end{subfigure}
\caption{Same as Fig.~\ref{fig:pointA} but with Point C: $\alpha'=0.075$, 
$m_{\chi}=1.4$ GeV, $m_{A'}=0.4$ GeV, and $m_{h'}=1.4$ GeV. }
\label{fig:pointC}
\end{figure}

The panels (a) in Figs.~\ref{fig:pointA}, \ref{fig:pointB} and \ref{fig:pointC}
imply that the $H_T$ distribution in the Vector 
Model is more enhanced at low $H_T$, compared with the Chiral Model.  
Note that the emitted dark photons  are mainly transverse 
in the Vector Model, but can be both transverse and longitudinal in 
the Chiral Model. There is also additional contribution from dark Higgs bosons
in the Chiral Model,  which was not included in previous studies. 
The enhancement at low $H_T$ is then
understood, for the unbroken splitting $\chi_s\rightarrow \chi_s A'_T$ 
contains soft singularity, whereas $\chi_s\rightarrow \chi_{-s} A'_L$ 
and $\chi_s\rightarrow \chi_{-s} h'$ do 
not.   This is the major feature that differentiates the chirality 
of the DM fermion. In particular, this feature is most useful, as the 
Yukawa coupling, characterized by the ratio $\frac{m_{\chi}}{m_{A'}}$,
is comparable to the gauge coupling\footnote{We have confirmed 
that the $H_T$ distributions from the Chiral Model with $Q_L=1$ and $Q_R=0$ and 
from the Vector Model with $Q_L=Q_R=1$ are exactly the same after 
proper normalization, when the Yukawa coupling is zero or negligible 
compared to $\alpha'$.}.  As exhibited in 
the panels (a) of Figs.~\ref{fig:pointA}, \ref{fig:pointB} and \ref{fig:pointC},  
the Chiral Model and the Vector Model are clearly distinguished 
for $\frac{m_{\chi}}{m_{A'}}=3.5$ (Point C) and  
$\frac{m_{\chi}}{m_{A'}}=2.5$ (Point B). For 
$\frac{m_{\chi}}{m_{A'}}=1.75$ (Point A), the distinction becomes less
obvious at large $H_T$, but is still significant at low $H_T$.

The $n_{A'}$ distribution is plotted in the panels (c) of
Figs.~\ref{fig:pointA}, \ref{fig:pointB} and \ref{fig:pointC},
in which the peak height in the $n_{A'}$ distribution is generally 
larger, while the peak $n_{A'}$ itself is lower, in the Vector Model 
than in the Chiral Model. This difference is again attributed to 
the additional emissions of longitudinally polarized dark
photons and dark Higgs bosons in the Chiral Model, which
increase the dark photon number. 

We point out that the dark Higgs boson appears only in the Chiral Model. It
can have important effects on the  patterns of the above observables, 
depending on the relation of the dark Higgs mass $m_{h'}$ 
to masses of the other particles in the model. If $m_{h'}$ is much 
larger than both $m_{\chi}$ and $m_{A'}$, the dark Higgs boson 
does not contribute to the dark shower, corresponding to the
curves labelled by ``Chiral model with T+L'' 
in the plots. As $m_{h'}$ is comparable to $m_{\chi}$ and $m_{A'}$, 
every dark Higgs boson produced in the shower accounts for two dark photons, 
altering the signals of lepton jets.  This case corresponds to the curves
labelled by ``Chiral model with T+L+h''. It is found that
the dark Higgs boson emission further pushes the distributions of the dark 
photon number to larger $n_{A'}$ in the Chiral Model, as indicated in 
the panels (c) of Figs.~\ref{fig:pointA}, \ref{fig:pointB} and \ref{fig:pointC}. 
At last, we observe in the panels~(b) and (d) that
the effects from the various next-to-leading-power, i.e., broken splittings 
are generally too small to be identified in the distributions.

We have emphasized the differences between our treatment of the 
dark shower and the splitting functions and that in Ref.~\cite{Zhang:2016sll}  
at the end of Sec.~\ref{subsec:splitting}.  The results of the Chiral Model 
using the program and the splitting functions in Ref.~\cite{Zhang:2016sll} 
correspond to the curves labelled by ``chiral model from Zhang
et al.''. Regardless of the general agreement, we cannot accommodate
some distinctions from those in Ref.~\cite{Zhang:2016sll}, which might be 
due to the different settings in the shower program and {\tt Pythia}.  
We have cross checked  our program with that of Ref.~\cite{Buschmann:2015awa} 
for the Vector Model, and confirmed  agreement on the average 
dark photon number.


\begin{figure}
\begin{subfigure}{0.49\textwidth}
\centering
\includegraphics[width=\textwidth]{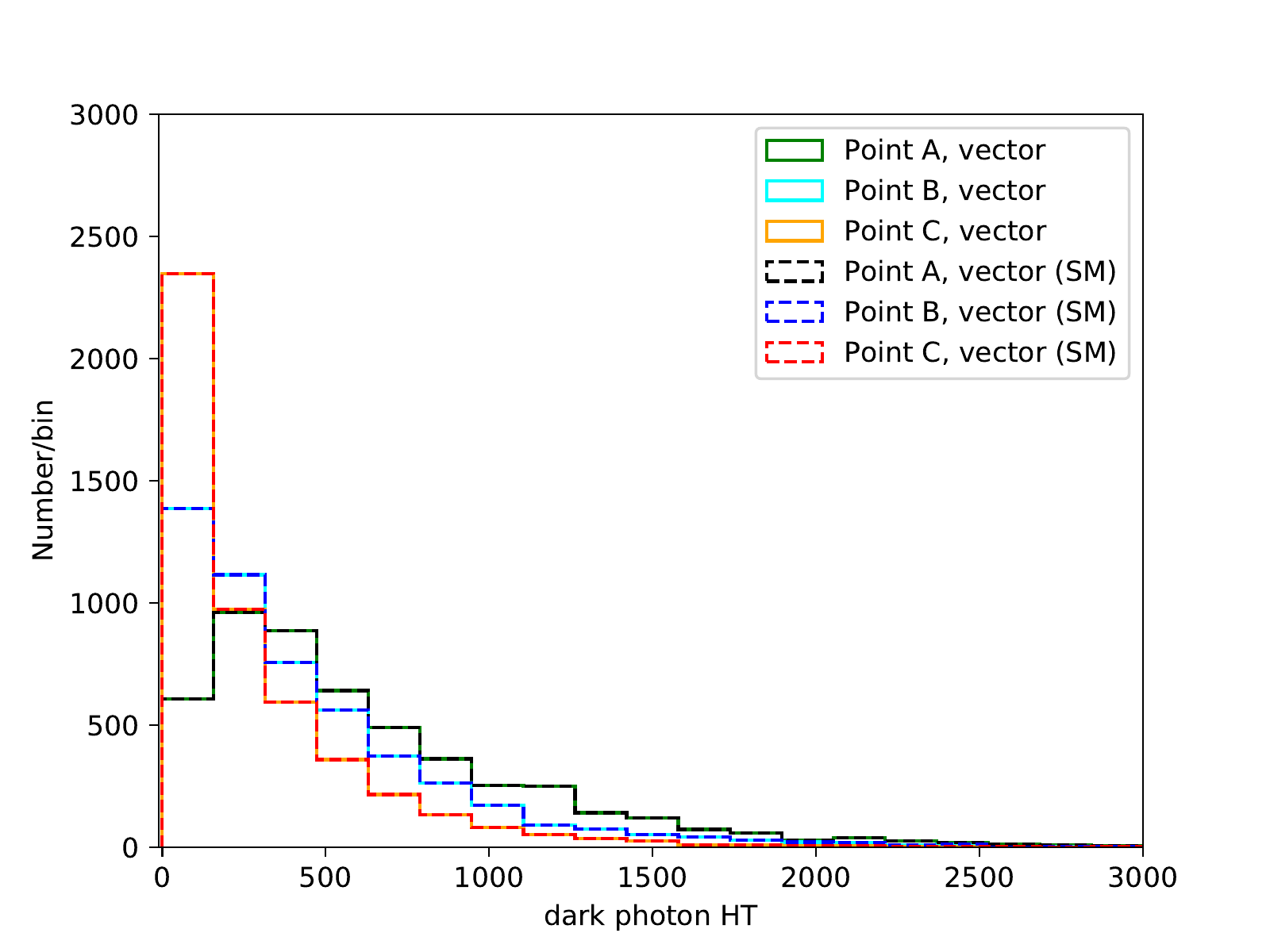}
\end{subfigure}
\begin{subfigure}{0.49\textwidth}
\centering
\includegraphics[width=\textwidth]{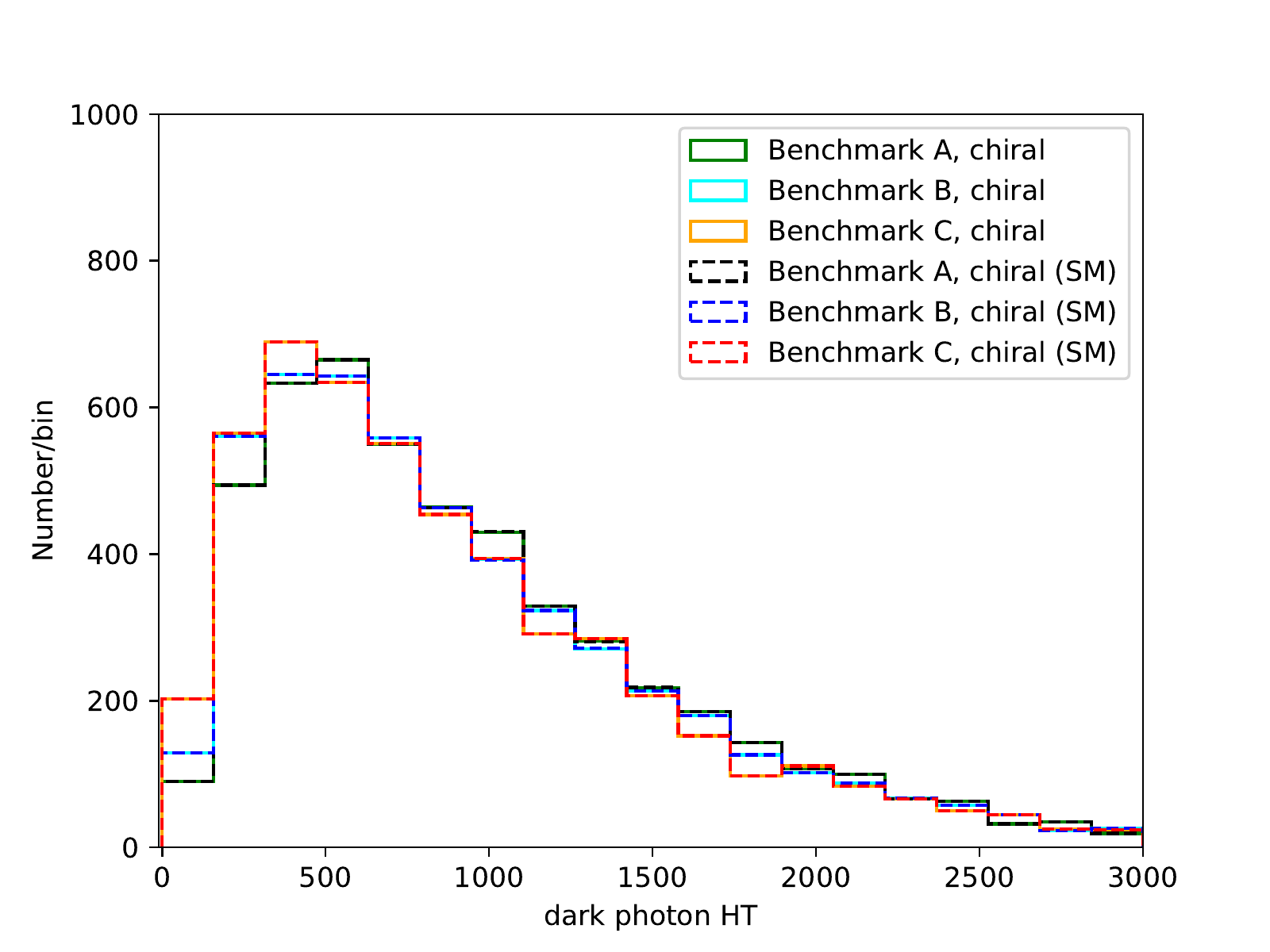}
\end{subfigure}
\caption{Comparison between $H_T$ with the dark photons as final states  
and that with the SM particles which the dark photons decay to as final states. }
\label{fig:SM}
\end{figure}

Although we considered the dark photons as final states in generating
Figs.~\ref{fig:pointA}, \ref{fig:pointB} and \ref{fig:pointC}, the results 
for the dark photon number $n_A$ and for the scalar sum $H_T$ of the dark photon 
transverse momenta are basically identical, when the SM particles which the dark 
photons decay to are taken as final states. The reason is that almost all dark 
photons decay to SM particles before they reach the detectors at LHC with 
the kinematic mixing chosen in this work, as stated before.
Moreover, the DM fermion produced in the hard process is highly boosted, 
so that all the particles in the dark shower are collimated, and contribute to
the above observables. We illustrate this fact by presenting the plots with 
both the dark photons and the SM particles as final states in Fig.~\ref{fig:SM}.

\begin{figure}
\begin{subfigure}{0.49\textwidth}
\centering
\includegraphics[width=\textwidth]{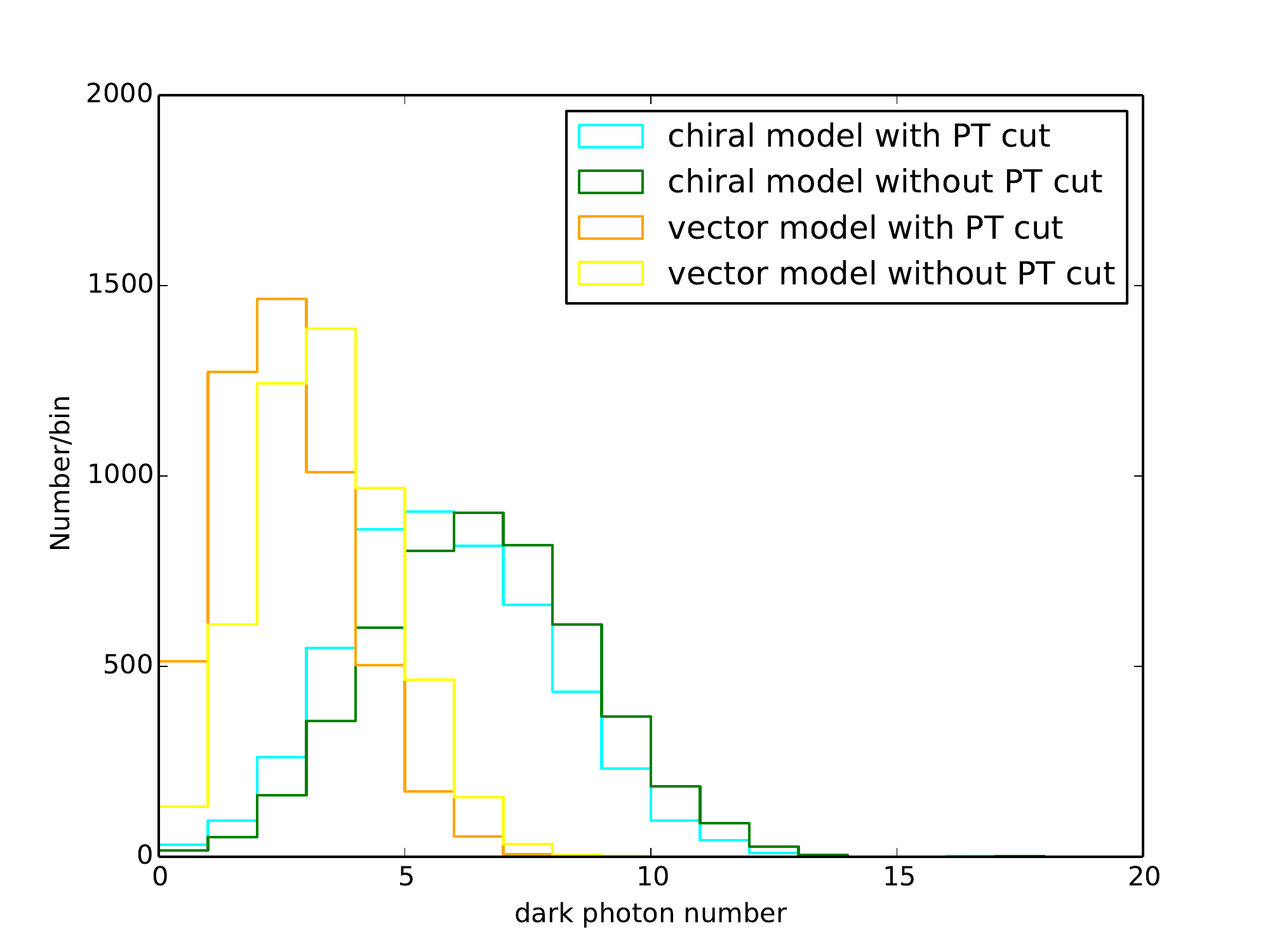}
\end{subfigure}
\begin{subfigure}{0.49\textwidth}
\centering
\includegraphics[width=\textwidth]{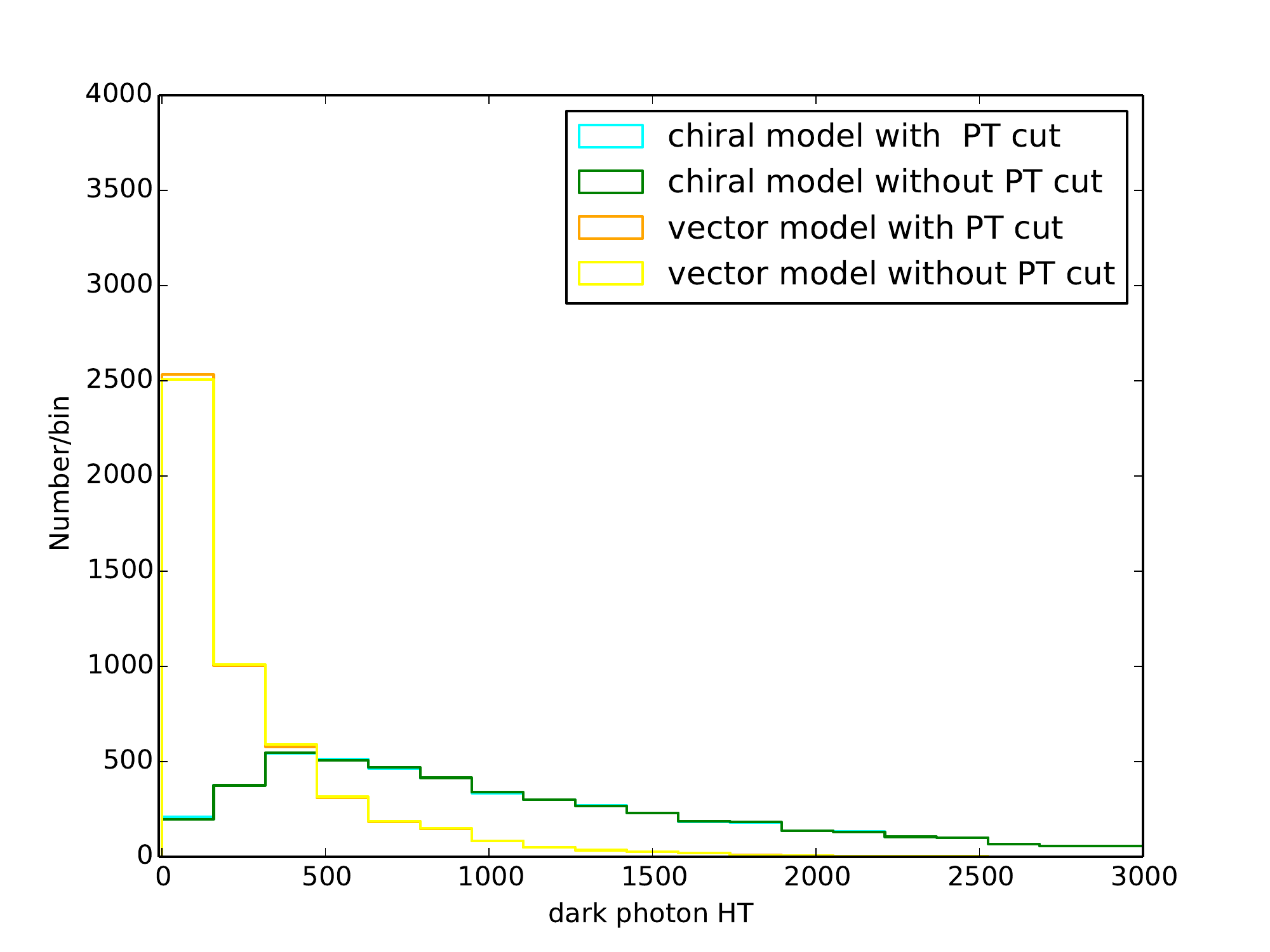}
\end{subfigure}
\caption{Distributions of dark photons without a cut and 
with a cut $p_T>10$ GeV for the Vector Model and the Chiral Model. 
The parameters are set to Point C.  }
\label{fig:cut}
\end{figure}

We then examine how the distributions of dark photons are 
affected by the cut imposed on the dark photon transverse momentum $p_T$ 
relative to the hard DM fermion. We plot in Fig.~\ref{fig:cut}
the $n_A$ and $H_T$ distributions with and without the cut $p_T>10$ GeV 
for Point C. It is found that the $H_T$ distribution is not modified 
by the cut, whereas the $n_A$ distribution exhibits a dependence on the 
cut. It confirms the expectation that the number of dark photons is not 
an infrared-safe observable and contains an inherent theoretical uncertainty. 
Nevertheless, the difference between the Chiral and the Vector Model is not 
washed out after imposing the $p_T$ cut, because the distributions shift 
along the same direction.

\begin{figure}
\centering
\includegraphics[width=0.45\textwidth]{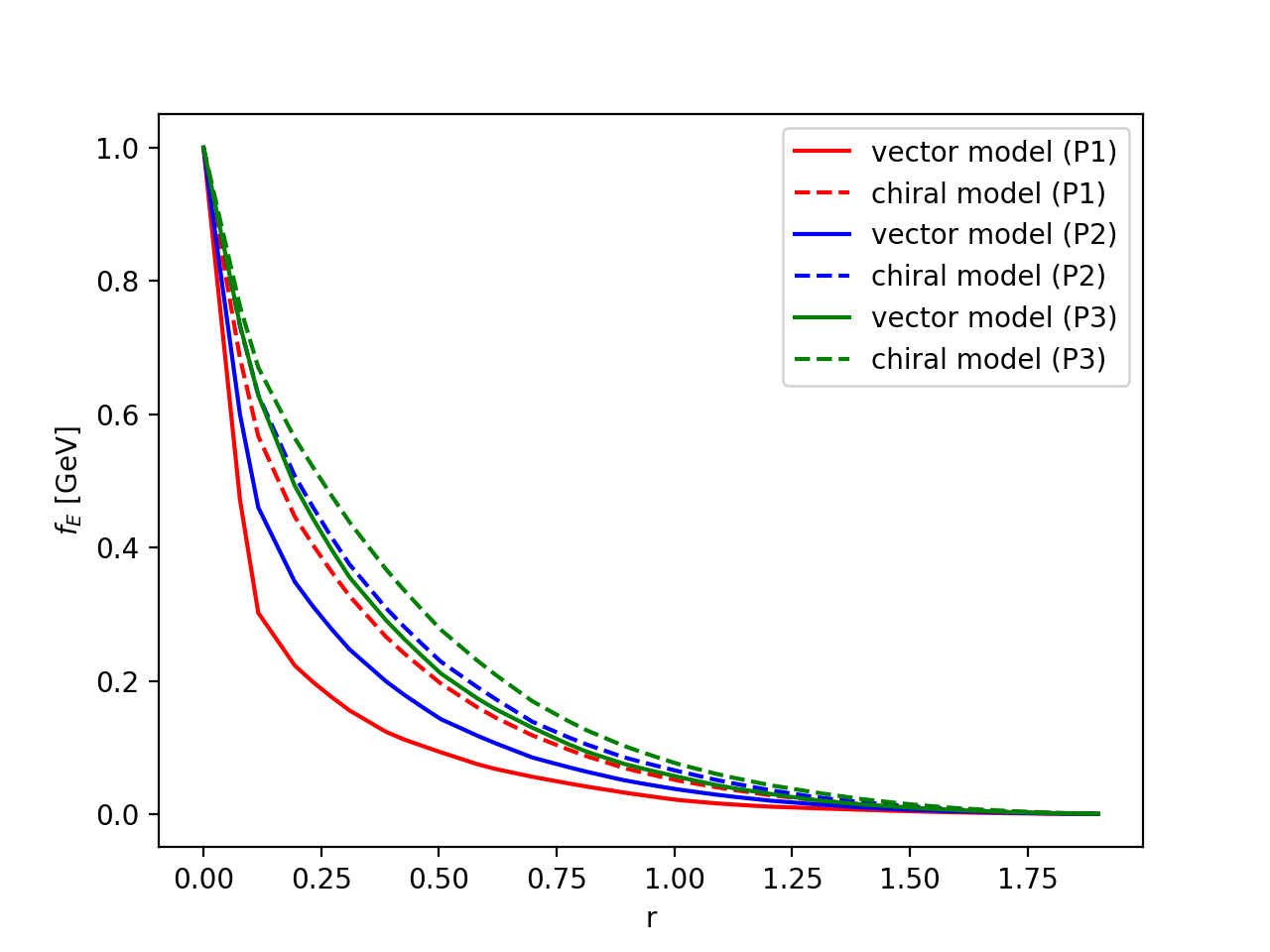}  
\includegraphics[width=0.45\textwidth]{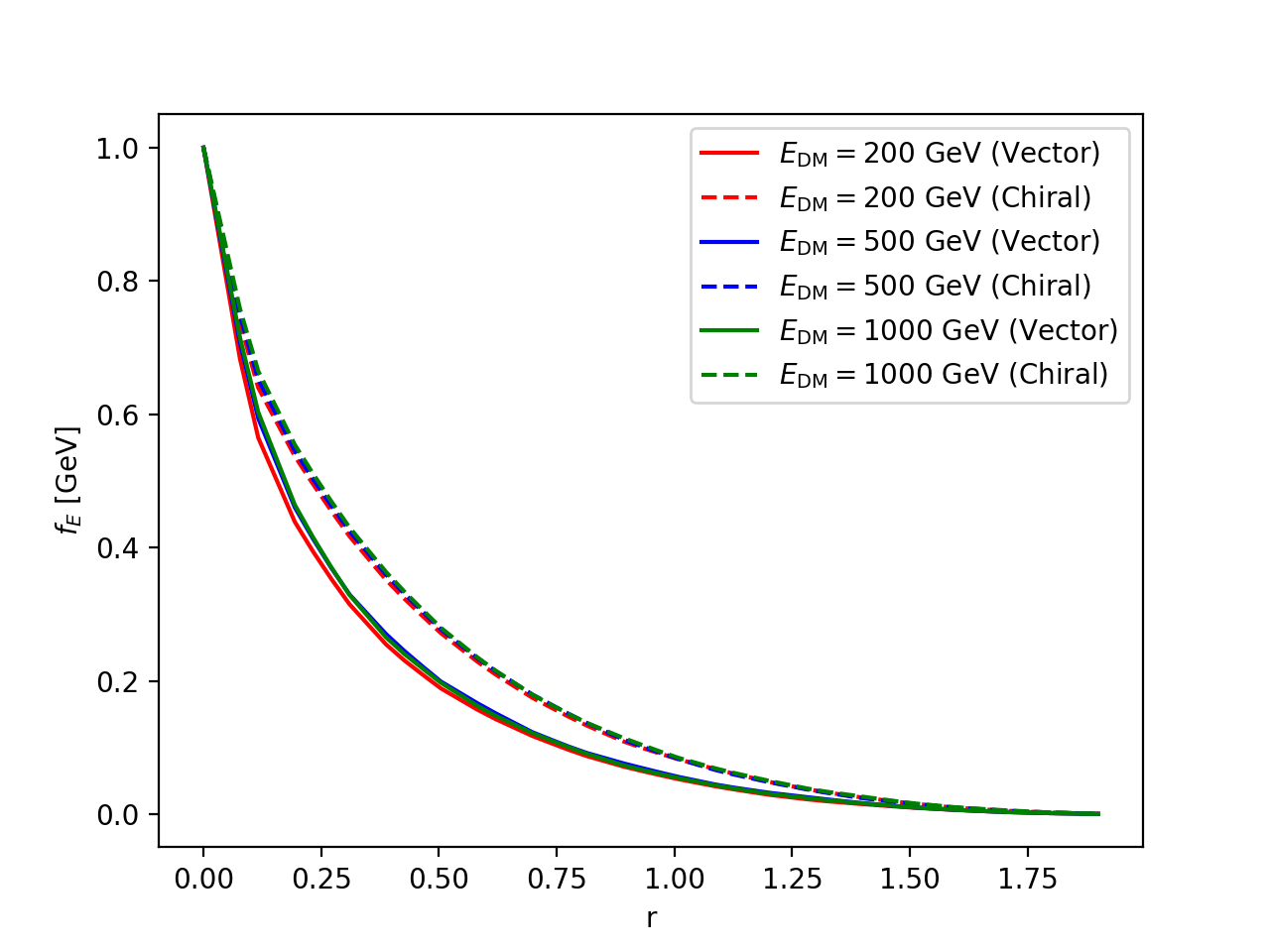}  
\caption{Left panel: energy profiles of DM jets for three benchmark points 
in the mono-jet channel ($p_T(j)>200$ GeV) at 14 TeV LHC.  Right panel:  
energy profiles of DM jets for Point A with different DM energies. }
\label{fig:lepjet}
\end{figure}

To examine the shape of DM jets for each benchmark point, we cluster
the final state particles radiated by a DM fermion using the anti-$k_t$ jet 
algorithm for the jet radius $R=2$ to determine the jet axis.  
We then average the energy deposit over $10^4$ DM jet events with 
respect to the distance to the jet axis. The jet profile is then 
described by the variable $f_E(r)$, defined as the energy fraction 
outside the cone with the radius $r<R$. The distributions of $f_E(r)$ 
from the Vector and Chiral Models for the three benchmark points are 
displayed in the left panel of Fig.~\ref{fig:lepjet}, which descend from 
$f_E(r=0)=1$ to $f_E(r=R)=0$ following different curves.  We notice 
that the jets are broader in the Chiral Model than in the Vector Model,
since longitudinally polarized dark photons and dark
Higgs bosons without the soft singularity in the momentum fraction $z$
can attain larger transverse momentum $k_T$ compared with 
transversely polarized dark photons, according to the Sudakov 
form factor in Eq.~(\ref{eq:sudakov}).
In the right panel, we exhibit the jet profile for the point A with 
different DM energies. It is seen that the jet profile is mainly 
determined by the DM fermion chirality, and almost independent of the DM energy.
This observation can be understood via the resummation formalism for the
jet energy profile~\cite{Li:2011hy}, whose behavior is mainly determined by 
the $r$-dependent and energy-independent double logarithm.  
It implies that the jet profile is an appropriate observable for 
differentiating the DM fermion chirality.

\section{Conclusion}

In this paper we  investigated the dark shower patterns generated 
by  energetic light DM fermions with different interactions to 
the dark photons at the LHC, evaluating the three observables
explicitly, the scalar sum of dark photon transverse momenta,
the dark photon number, and the energy profile of DM jets.
Our work was motivated by the connection of the DM chiral property under a 
dark $U(1)$ gauge group to the mass origin of the dark sector, which 
could be realized at least in the simple Chiral and Vector Models considered here.
It was shown that the DM chirality can indeed be distinguished by measuring 
the dark shower patterns: the shower is dominated by soft transversely
polarized dark photons in the Vector Model, while it contains extra
energetic longitudinally polarized dark photons and dark Higgs bosons in
the Chiral Model. Especially, the jet energy profile, mainly determined 
by the DM fermion chirality and almost independent of the DM energy, 
seems to be an appropriate observable for the purpose.  

Compared with the literatures on 
this subject, we have derived the complete set of $1\rightarrow 2$ 
splitting functions with the DM fermion as the initial state in the 
``DM fermion+dark $U(1)$'' scenario. Based on  these splitting functions, 
our implementation of the dark shower exhibits several novelties, making 
the analysis more accurate and valuable:

1.\ We specified the helicities of the DM fermions in the splitting
functions and stressed that this specification is important for the 
Chiral Model, especially when the Yukawa coupling is comparable to the 
dark gauge coupling. 

2.\ We analyzed the effects of the dark Higgs boson in different
limits of the dark Higgs mass. 

3.\ We included the symmetry breaking effects in the dark shower through
a class of new splitting functions at power of $\frac{m^2}{k_T^2}$, 
though their effects on the shower patterns were found to be minor in general. 

With the framework being solidly built up for correctly modeling the
dark shower phenomena, we plan to carry out a careful collider analysis 
and related searching strategies in the forthcoming paper. 
It is also obvious that our formalism can be applied to more complicated 
and realistic models, and extended to include splittings
of other initial particles, such as dark Higgs bosons, dark photons, 
etc..\.   We will address these subjects in future publications.

\

\

\


\noindent \textbf{Acknowledgements} We thank the discussions with Tao Han,  
Myeonghun Park, Brock Tweedie and Mengchao Zhang.  This work was  partly 
supported in part by National Research Foundation of Korea (NRF) Research
Grant NRF-2015R1A2A1A05001869 (PK, TM), and by the Ministry
of Science and Technology of R.O.C. under Grant No. MOST-104-2112-M-001-037-MY3.

\appendix
\section{Examples of Splitting Function Calculation}
\label{sec:example}

\begin{figure}[t]
\centering
\includegraphics[width=7cm]{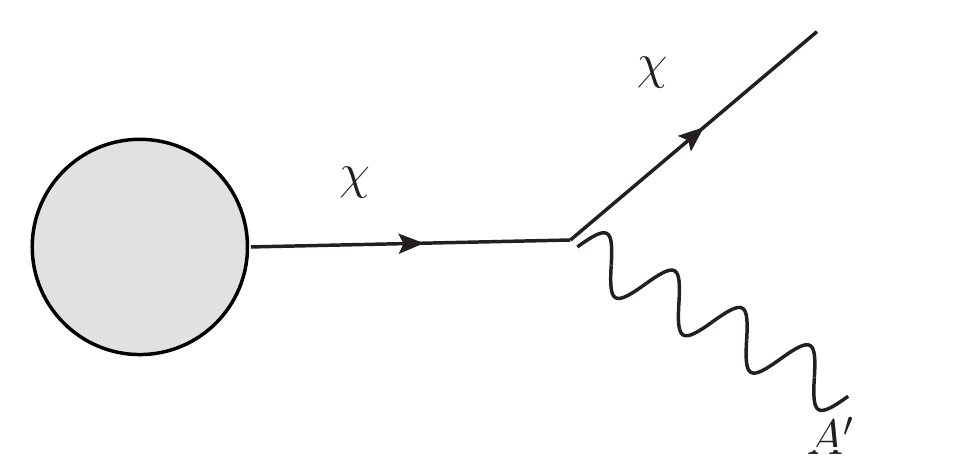}
\caption{The $\chi\rightarrow \chi A'$ splitting function.}
\label{fig:amplitude_long}
\end{figure}

We take the processes $\chi_{s}\rightarrow \chi_{-s} A'_L$ and $\chi_{s}\rightarrow
\chi_{s} A'_L$ in Fig.~\ref{fig:amplitude_long} as examples to
demonstrate how to calculate the splitting functions for $A\rightarrow B+C$.   
We follow the methods in Ref.~\cite{Chen:2016wkt} basically  by 
imposing the GEG, in which the amplitudes
involving longitudinal vector bosons are derived by summing over
both the Goldstone components and the remnant gauge components.  
To evaluate the collinear  splitting amplitudes and  the
splitting functions,   the total amplitude 
for a physical process should be factorized  into the form
\be
\label{eq:fac}
i{\mathcal M}=i{\mathcal M}_{split}\frac{i}{k_A^2-m_A^2}i{\mathcal M}_{hard} 
+ \text{power\  suppressed}.
\ee
\noindent The collinear splitting function is then related to the splitting 
amplitude $i{\mathcal M}_{split}$ via 
\be\label{eq:split_fun}
\frac{d{\mathcal P} }{dzdk_T^2}= \frac{1}{16\pi^2} z\zb 
\frac{\overline{|{\mathcal M}_{split}|^2}}{\tilde{k}_T^4}.
\ee
To satisfy the factorization condition in Eq.~(\ref{eq:fac}),  we need to 
write the fermion propagator of the initial virtual state as
\be
\frac{\slashed{k}+m_{\chi}}{k^2-m_{\chi}^2} =
\frac{\sum_su^0_s(k)\bar{u}^0_s(k)}{k^2-m_{\chi}^2} + {\cal O}\left(\frac{1}{E}\right),
\ee
\noindent with $u^0_s(k)$ being the ``on-shell'' wave functions,
\[ u^0_-(k)=\sqrt{E+|\vec{k}|}\left( {\bray{c}  \xi_- \\
                        \frac{m_{\chi}}{ E+|\vec{k}|} \xi_- \eray}\right)
 \hspace{1cm} u^0_+(k)=\sqrt{E+|\vec{k}|} \left( {\bray{c} \frac{m_{\chi}}{ E+|\vec{k}|} \xi_+ \\
                          \xi_+ \eray}\right).
\]
\noindent  The factorization form makes clear that only 
the ``on-shell'' wave functions  contribute nontrivially to the 
splitting amplitude $i{\mathcal M}_{split}$ and then to the collinear 
splitting function.  


We now compute the amplitude for $\chi_{s_1}\rightarrow \chi_{s_2}A_L'$,  
\be
i{\mathcal M}_{\chi_{s_1}\rightarrow \chi_{s_2} A_L'} 
= i{\mathcal M}_{\chi_{s_1}\rightarrow \chi_{s_2} \phi'} + \text{phase}\cdot  i{\mathcal M}_{\chi_{s_1}\rightarrow \chi_{s_2} A_n'},
\ee
where the relative phase between the two amplitudes 
${\mathcal M}_{\chi_{s_1}\rightarrow \chi_{s_2} \phi'}$ and 
${\mathcal M}_{\chi_{s_1}\rightarrow \chi_{s_2} A_n'}$ can be 
obtained in the same way as Eq.~(B16) in Ref.~\cite{Chen:2016wkt}. 
We define the covariant derivative 
$D_{\mu}\Phi'=(\partial_{\mu}-ig'Q_{\Phi'}A_{\mu}')\Phi'$ with $g'Q_{\Phi'}>0$,
by means of which the mixing Lagrangian becomes 
$-m_{A'}\partial_{\mu}\phi' A^{'\mu}$ with a minus sign. The phase is 
then given by
\be
\text{phase} =
\left\{
\begin{array}{rcl}
-i & \ \ \ & \text{for  incoming momentum}, \n \\
i  &\ \ \ & \text{for outgoing momentum}. \n
\end{array}
\right.
\ee
%




We specify the helicities, and divide the splittings into the 
helicity-flipping one $\chi_{s}\rightarrow \chi_{-s} A'_L$~(leading power) 
and the helicity-conserving one $\chi_{s}\rightarrow \chi_{s} A'_L$~(next-to-leading power). 
The $\chi_{s}\rightarrow \chi_{-s} A'_L$ splitting amplitude is written as
\be
i{\mathcal M}^{\chi_{s}\rightarrow \chi_{-s}A'_L}_{split}=
i\sqrt{2}g'\frac{m_{\chi}}{m_{A'}}Q_{\Phi'}
\bar{u}^0_{-s}(k_B)\gamma_5u^0_{s}(k_A) + {\cal O}\left(\frac{m}{E}\right),
\ee
\noindent 
in which the power suppressed term comes from the gauge component contribution.
The Goldstone component leads to 
\be
i{\mathcal M}^{\chi_{s}\rightarrow \chi_{-s}A'_L}_{split}= 
i\sqrt{2}g'\frac{m_{\chi}}{m_{A'}}Q_{\Phi'} \frac{1}{\sqrt{2\zb}}k_T,
\ee
via which we obtain, according to Eq.~(\ref{eq:split_fun}), the splitting function 
for $\chi_{s}\rightarrow \chi_{-s}A'_L$ given in Eq.~(\ref{eq:split_unbroken}).
%
%

The $\chi_{s}\rightarrow \chi_{s} A'_L$ splitting amplitude is decomposed into 
\be
i{\mathcal M}^{\chi_{s}\rightarrow \chi_{s}A'_L}_{split} 
&=& i{\mathcal M}^{\chi_{s}\rightarrow \chi_{s}A'_n}_{split} 
+ \text{phase}\cdot  i{\mathcal M}^{\chi_{s}\rightarrow \chi_{s}\phi'}_{split} , 
\ee
\noindent with 
\be
 i{\mathcal M}^{\chi_{s}\rightarrow \chi_{s}A'_n}_{split} 
 &=& ig'\sum_{s=L,R}Q_s\bar{u}^0_{s}(k_B)\gamma^{\mu}P_{s}u^0_{s}(k_A)\epsilon_{\mu}, \n \\ 
 i{\mathcal M}^{\chi_{s}\rightarrow \chi_{s}\phi'}_{split} 
 &=& i\sqrt{2}g'\frac{m_{\chi}}{m_{A'}}Q_{\Phi} \bar{u}^0_{s}(k_B)\gamma_5u^0_{s}(k_A),
\ee
%
%
\noindent 
where $P_{s}$ is the operator to project out the left-handed chirality 
($P_L=\frac{1-\gamma_5}{2}$) or the right-handed chirality 
$P_R=\frac{1+\gamma_5}{2}$. A straightforward derivation yields the 
splitting amplitudes 
\be
i{\mathcal M}^{\chi_{L}\rightarrow \chi_{L}A'_L}_{split} &=& 
ig'm_{A'}\frac{2}{z\sqrt{\zb}}\left(Q_L\zb+\frac{z^2m_{\chi}^2}
{2m_{A'}^2}(Q_L-Q_R)\right), \n \\
  i{\mathcal M}^{\chi_{R}\rightarrow \chi_{R}A'_L}_{split} &=& 
  ig'm_{A'}\frac{2}{z\sqrt{\zb}}\left(Q_R\zb-\frac{z^2m_{\chi}^2}
  {2m_{A'}^2}(Q_L-Q_R)\right). \n \ee
\noindent 
Combining  the $s=R(+\frac{1}{2})$ and $s=L(-\frac{1}{2})$ pieces, we have 
\be
i{\mathcal M}^{\chi_{s}\rightarrow \chi_{s}A'_L}_{split} &=& 
ig'm_{A'}\frac{1}{z\sqrt{\zb}}\left(2Q_s\zb+(-1)^{s+\frac{1}{2}}
\frac{z^2m_{\chi}^2}{m_{A'}^2}Q_{\Phi'}\right).
\ee
\noindent 
Inserting the above expression into Eq.~(\ref{eq:split_fun}) leads to
the splitting function for $\chi_s\rightarrow \chi_s A'_L$  
in Eq.~(\ref{eq:split_broken}),
\be
\frac{d{\mathcal P} }{dzdk_T^2}(\chi_s\rightarrow \chi_s A'_L)= 
\frac{\alpha'}{2\pi} \frac{1}{2z}\left(2Q_s\zb+(-1)^{s+\frac{1}{2}}
\frac{z^2m_{\chi}^2}{m_{A'}^2}Q_{\Phi'}\right)^2 \frac{m_{A'}^2}{\tilde{k}_T^4}.
\ee

\end{document}